\documentclass[preprint]{aastex}

\shorttitle{New analysis of NGC 1866}
\shortauthors{V. Testa et al.}

\begin{document}

\title{The Large Magellanic Cloud globular cluster NGC 1866: new data, 
new models, new analysis\footnote{Based on observations obtained at ESO, 
La Silla, Chile}}

\author{Vincenzo Testa}
\affil{Osservatorio Astronomico di Roma}
\affil{Via Frascati, 33 00040 Monteporzio Catone, Italy}
\email{testa@oar.mporzio.astro.it}

\author{Francesco R. Ferraro\altaffilmark{1}}
\affil{European Southern Observatory}
\affil{Karl Schwarzschild Strasse 2,D-85748 Garching bei M\"unchen, Germany}

\author{Alessandro Chieffi}
\affil{Istituto di Astrofisica Spaziale, C.N.R.}
\affil{Via del Fosso del Cavaliere, 00100 Roma, Italy}

\author{Oscar Straniero}
\affil{Osservatorio Astronomico di Collurania}
\affil{Via Maggini, 88 64100 Teramo, Italy}

\author{Marco Limongi}
\affil{Osservatorio Astronomico di Roma}
\affil{Via Frascati, 33 00040 Monteporzio Catone, Italy}
\and
\author{Flavio Fusi Pecci\altaffilmark{1}}
\affil{Stazione Astronomica di Cagliari}
\affil{09012 Capoterra, Italy}

\altaffiltext{1}{On leave from Osservatorio Astronomico di Bologna, via Ranzani 1, I-40126 Bologna, ITALY}

\begin{abstract}
\scriptsize
We present a new deep (down to $V \sim 24$) photometry of a 
wide region ($\sim 6'\times 6'$) around the LMC globular cluster NGC1866: our 
sample is much larger (by more than a factor three) than any previous 
photometry and with a main sequence which may be considered
complete, down to at least 3 mag below the 
brightest MS star: such an occurrence allows a meaningful and robust 
comparison with various theoretical scenarios produced by means of models
computed with the evolutionary code FRANEC.
Both age and present mass function slope, $\alpha$, are derived 
by a fit to the available MS and by the use of the parameter 
$\Delta\sigma$, which is simply the difference, in $\sigma$'s, between the 
observed and predicted integrated MS luminosity functions. 
Our main conclusions are:

a) the adoption of standard models (i.e. computed by adopting the Schwarzschild
criterion to fix the border of the convective core) allows a fair fit to the 
MS for an age of the order of 100-140 Myr and a present mass function having
a slope $\alpha$ between 2.3 and 1.9, the exact values depending on the 
adopted distance modulus. It is moreover possible to reproduce the 
average He clump luminosity while the total number of stars predicted in the 
He clump is twice the observed value: this means that we re-obtain and confirm 
the first finding of \cite{bec83}, according to whom the simple adoption of a 
"classical" scenario leads to a neat discrepancy concerning the prediction 
of the number of stars in the He clump. 

b) the adoption of models computed by increasing the size of the convective 
core by a certain amount, i.e. $0.25~H_p$, leads to a fair fit to the main 
sequence only for a visual distance modulus $(m-M)_V\simeq18.6$, an age  
$t \simeq 200~ Myr$ and $\alpha\simeq2.2$. In this case, the total number
of He clump stars is well reproduced, although the luminosity function of the 
He clump itself is predicted to be sistematically less luminous than observed.

The previous conclusions are based on the assumption that there is no
appreciable population of binaries in NGC 1866. Though there are not yet 
sufficient data on the frequency of binary systems in these clusters, we
analyzed how the previous scenarios would change if a consistent 
$(\simeq 30\%)$ population of binary systems were present in the cluster. 
This choice is based on the fact that a fraction of binaries
of the order of 30\% has already been found in NGC 1818, a cluster similar 
to NGC 1866 \citep{els98}. The inclusion of a 30\% binary population leads to 
the following conclusions:

c) the adoption of the standard models now leads to a good fit to the entire 
luminosity function, i.e. main sequence, turn off, and He clump stars, for 
a visual distance modulus $(m-M)_V = 18.8$, an age  $ t \simeq 100$ 
Myr and a mass function slope  $\alpha \simeq 2.4$, thus largely removing 
the ``classical'' discrepancy between observed and predicted number of 
stars in the He burning clump. The quoted visual distance modulus
constrains the unreddened distance modulus 
$(m-M)_0$ within 18.50 and 18.62, depending on the reddening (whose
most common values available in the literature range from 0.05 to 0.10).

d) at variance with point c), the fit obtained by using models computed with 
an enlarged convective core gets worse when a binary component is taken into 
account. This is due to the fact that the presence of binary systems 
increases the existing discrepancy between the observed and predicted 
clump luminosity, since the He clump is predicted to be even less luminous 
than in absence of binaries.

As a consequence of this  analysis, we think that 
the next step towards a proper understanding of NGC 1866, and similar 
clusters, must include the accurate determination
of the frequency of binary systems that will be hopefully performed with
the incoming Cycle 8 HST observations of NGC~1866.

\end{abstract}


\section{Introduction}

NGC 1866 is one of the most largely studied stellar clusters in the Large Magellanic
Cloud (LMC). Since the first photometric study by \cite{arp67}, who discovered
a large number of member Cepheid variables, this cluster has been recognized
as an ideal laboratory for testing stellar evolutionary models of 
intermediate-mass stars. NGC 1866 is one of the most populous LMC clusters, 
having a mass comparable to that of a typical globular cluster in the Galaxy 
\citep[$M=1.25\pm 0.25\times 10^{5}M_{\odot}$][]{fis92} and is relatively 
young to study the evolution of intermediate-mass ($M \sim 5M_{\odot }$)
stars. These stars develop a convective core during the H-burning phase and 
represent the ideal target for ana analysis of the real extension of the 
convective cores, i.e., to see whether the size of convective cores, 
as determined by the 
Schwarzschild criterion, to which we will refer as ``standard'', must be 
``artificially'' increased  in order to fit the observational data.
This is a long-standing problem whose solution would deeply influence many 
aspects of our present understanding of the universe: in fact it would fix a 
key parameter in the computation of the stellar evolution which would reflect,
 e.g., on the intermediate age clusters,  on the 
determination of the various mass limits, like $m_{up}$ and $m_{up}^{'}$ 
\citep[see][]{cgr71,bec79,ccp85} and on the initial-final mass relation, 
on the relation between He-core mass and initial mass for massive stars 
that is linked to the problem of metal enrichment.

NGC 1866 has often been considered a "proof" for the existence of the 
convective overshooting \citep{bec83,chi89}. 
Actually the fit to this (or another) cluster can only give us information 
on the likely size of the convective core but it ascertain
the mechanism which determines the size. For 
reasons sinking in the history of the computation of stellar models in 
the range of massive stars (whose analysis goes well beyond the
purposes of this paper), this problem has always been (unfortunately!),
interpreted in terms of ``overshooting'', therefore as a proof 
of the existence or non=existence of a specific phenomenon where 
the convective cells cross the classical border of the 
convective core and mix more matter than predicted in the ``classical'' 
scenario, thus producing a larger amount of fuel. This direct
connection is meaningless: in fact, if the observations showed that the 
convective core must be larger than currently predicted by the models, 
various causes could be considered. Just to mention two 
possibilities other than the``overshooting'', the simple change from 
the old Cox-Tabor opacities \citep{cox76} to the Los Alamos ones \citep{hue77},
 which occurred at the beginning of the 80's, led, by itself, to a 
significant increase of the classical size of the convective core. A second 
possibility, proposed for example by A. Maeder (private communication) is 
a rotationally induced mixing that leads to a more massive convective core.

\cite{bec83} were the first to point out
that the ``standard'' models could not fit the observed 
ratio of main sequence vs He clump stars $(N_{MS}/N_{He})$: they found that 
their standard models predicted at least 4 times more giant stars
than observed with respect to the main sequence ones. In the following years,
\citet[hereafter C89]{chi89} and \citet[hereafter B89]{bro89}, studied in 
detail the properties of the stellar content of NGC 1866, using CCD data. 
A detailed discussion of the general problem as well as the results of C89
and B89 are given in \cite{woo91}. Briefly, B89 concluded that the standard 
models reproduce both the color-magnitude diagram (CMD) morphology and the 
luminosity function (LF) of the cluster, while C89 concluded that models 
including a certain amount of overshooting were required.

The problem has been  further discussed  in later papers 
\citep[hereafter BCP94]{lat91,bro94} without reaching an agreement.
A larger sample of data led BCP94 to conclude that their previous set of data 
was substantially correct, whereas the ones adopted by C89 were affected by 
stochastic fluctuations. Actually, the sample adopted
by C89 included 554 MS stars (down to $V=19.2$) and 41 He burning giants, 
while the latest set of data presented by BCP94 included 
(down to the same magnitude) 462 MS stars and 65 giants. All these numbers 
were already corrected by the authors for completeness, crowding and field 
stars contamination.

In order to contribute to clear up the situation, we began a project 
specifically devoted to a complete photometric study of the stellar 
population in NGC 1866, including a substantial part of the MS. In this paper,
we present a new deep (down to $V \sim 24$) photometry of a wide region 
$\sim 6'\times 6'$ around the cluster.

The observations and reduction procedures are described in sect. 2, while
the morphology of the inferred CMD is presented and discussed in sect. 3. 
Sect. 4 is devoted to the discussion of the derived LF, with a special 
attention to the field decontamination and the completeness estimate and 
correction, together with the comparison with previous studies. The comparison 
with various theoretical scenarios is discussed in sect. 5. A final section 
with summary and conclusions follows.

\begin{deluxetable}{lcc}
\small
\tablecaption{Observed Frames\label{obs}}
\tablewidth{0pt}
\tablehead{\colhead{Image} & \colhead{filter} & \colhead{exp. time}}
\startdata
 FA0122{*}& V & 30   \textit{ s}\\
 FA0124   & B & 60   \textit{ s}\\
 FA0125   & B & 60   \textit{ s}\\
 FA0127   & V & 15   \textit{ s}\\
 FA0128   & V & 15   \textit{ s}\\
 FA0130   & B & 60   \textit{ s}\\
 FA0131   & V & 5    \textit{ s}\\
 FA0133   & B & 15   \textit{ s}\\
 FA0134   & V & 480  \textit{ s}\\
 FA0136   & B & 120  \textit{ s}\\
 FA0139   & B & 1500 \textit{ s}\\
 FB0140   & V & 30   \textit{ s}\\
 FB0142   & B & 90   \textit{ s}\\
 FB0143   & B & 1500 \textit{ s}\\
 FB0144{*}& V & 480  \textit{ s}\\
\enddata
\end{deluxetable}

\section{Observations, Reductions and Calibrations}

\subsection{Observations and Reductions}

The data have been obtained at the 2.2m ESO-MPI telescope at La Silla (Chile),
on Jan. 27th, 1993. We used EFOSC2 equipped with the CCD \#19 Thomson of size  
 $1024 \times 1024$ pixels, with a scale of $0.363 \arcsec$/pix, which yields a
 total field of view of $\sim 6'\times 6'$. During the night we secured B,V 
frames in two fields around NGC 1866: $FA$, roughly centered on the cluster
center and $FB$, located at $\sim 6'$ SE from the cluster center and observed
for statistical field decontamination purposes.

Figure \ref{fig1} shows a computer map of the two fields. A circle is drawn at
$ \sim 6'$ from the cluster center and encloses the region dominated by stars 
belonging to the cluster. According to \cite{ber92}, at larger distances the
contribution of the cluster stars to the total population of the field is 
expected to drop below 1\%. The average observational conditions were good, 
and the seeing was around $1\arcsec$. A journal of the observations is 
reported in Table \ref{obs}.

The reductions were performed using DAOPHOT \citep{ste87}, implemented on
IRAF\footnote{IRAF is operated by AURA, Inc., under cooperative agreement 
with the National Science Foundation}. In order to identify stars within each 
frame, we followed the standard procedure as described on the DAOPHOT manuals, 
and used the routine ALLSTAR to perform the final PSF-fit.

Each frame was reduced independently, then a frame in each filter has been 
chosen as \textit{reference} frame for coordinate and magnitude 
transformations (reference frames are marked with an asterisk in Table 
\ref{obs}); then all the frames have been transformed to the 
reference system in order to get a homogeneous set of instrumental 
magnitudes and coordinates. 

A set of averaged instrumental magnitudes (weighed according to the 
photometric quality of the frames) have then been obtained for each field. 
The overlap region between the two adjacent fields has been used to transform 
the coordinates into a common local system. In this arbitrary system, 
the center of the cluster has been estimated at pixel  (520,520)  in 
$FA$ (see Figure \ref{fig1}).

\subsection{Calibrations}

Since the observations were performed during slightly non-photometric 
conditions, we used the well-calibrated photometric sequence of standard 
stars obtained by \cite{wal95} in the surrounding region of NGC 1866, to 
link the instrumental magnitudes to the standard Johnson system. The 
cross-correlation between our data set and Walker's own identified 2294 stars 
in common, which were used to calibrate the final catalog.

The list of the final calibrated B,V magnitudes and position for 12680 ($FA$)
+ 3044 ($FB$) stars identified in our survey is available in electronic form.
Table \ref{clu} reports a sample of the table of photometry of $FA$ while 
Table \ref{fie}
reports the analogous for $FB$. Figure \ref{fig2}a shows the CMD for 
stars in $FA$ (hereafter referred to as CLUSTER sample); in the same way, 
Figure \ref{fig2}b shows the CMD for stars lying at $r>6'$ from the 
cluster center, which are assumed to be representative of the field 
population around the cluster (hereafter referred to as FIELD sample).

\begin{deluxetable}{lcccccc}
\small
\tablecaption{Photometry of CLUSTER frame\label{clu}}
\tablewidth{0pt}
\tablehead{\colhead{Id.num.} & \colhead{$X$} & \colhead{$Y$} & \colhead{$V$}
& \colhead{$\sigma(V)$} & \colhead{$B-V$} & \colhead{$\sigma(B-V)$}}
\startdata
    1 & 170.77 &   0.86&  19.68 &   0.01 &   0.01 &   0.03 \\
    2 & 859.86 &   1.49&  20.31 &   0.02 &   0.09 &   0.03 \\
    3 & 867.64 &   8.95&  22.23 &   0.07 &   0.45 &   0.10 \\
    4 & 528.30 &   1.72&  18.91 &   0.01 &  -0.05 &   0.01 \\
    5 & 466.28 &   1.87&  22.48 &   0.07 &   0.61 &   0.09 \\
    6 & 611.78 &   2.28&  22.78 &   0.09 &   0.32 &   0.13 \\
    7 &  57.47 &   2.49&  21.58 &   0.04 &   0.47 &   0.07 \\
    8 & 479.96 &   2.62&  20.72 &   0.01 &   0.10 &   0.02 \\
    9 & 942.88 &   2.92&  22.65 &   0.09 &   0.95 &   0.16 \\
   10 & 851.31 &   3.07&  21.45 &   0.05 &   0.24 &   0.06 \\
   11 & 368.49 &   3.28&  22.90 &   0.10 &   0.11 &   0.13 \\
   12 & 129.26 &   8.74&  21.33 &   0.02 &   0.21 &   0.04 \\
   13 &  25.95 &   3.77&  22.36 &   0.09 &   0.54 &   0.12 \\
   14 & 295.52 &   5.51&  21.49 &   0.03 &   0.35 &   0.04 \\
   15 & 444.89 &   5.91&  19.03 &   0.00 &  -0.07 &   0.01 \\
   16 & 104.19 &   5.99&  22.64 &   0.09 &   0.48 &   0.13 \\
   17 & 224.52 &   6.15&  22.27 &   0.04 &   0.47 &   0.08 \\
   18 & 319.41 &  11.00&  22.39 &   0.06 &   0.54 &   0.09 \\
   19 & 180.74 &   6.75&  23.20 &   0.13 &   0.73 &   0.20 \\
   20 & 383.21 &   9.18&  21.94 &   0.03 &   0.32 &   0.05 \\
\enddata
\end{deluxetable}

\begin{deluxetable}{lcccccc}
\small
\tablecaption{Photometry of FIELD frame\label{fie}}
\tablewidth{0pt}
\tablehead{\colhead{Id.num.} & \colhead{$X$} & \colhead{$Y$} & \colhead{$V$}
& \colhead{$\sigma(V)$} & \colhead{$B-V$} & \colhead{$\sigma(B-V)$}}
\startdata
    1 &   58.44 &  1.00 & 19.71 & 0.02 &  0.66 & 0.07 \\
    2 &  175.82 &  1.20 & 22.03 & 0.04 &  0.33 & 0.11 \\
    3 &  572.97 &  1.74 & 21.06 & 0.04 &  0.22 & 0.05 \\
    4 &    5.74 &  1.65 & 22.75 & 0.16 &  1.59 & 0.50 \\
    5 &  544.28 &  4.60 & 21.57 & 0.03 &  0.23 & 0.05 \\
    6 &  444.44 &  2.68 & 21.75 & 0.04 &  0.27 & 0.05 \\
    7 &  457.46 &  2.71 & 21.20 & 0.01 &  0.34 & 0.03 \\
    8 &  366.48 &  4.05 & 21.56 & 0.03 &  0.53 & 0.05 \\
    9 &  747.11 &  5.92 & 21.79 & 0.05 &  0.34 & 0.07 \\
   10 &  109.36 &  4.62 & 21.38 & 0.02 &  0.19 & 0.04 \\
   11 &  990.28 & 12.23 & 21.20 & 0.06 &  0.11 & 0.11 \\
   12 &  333.27 &  5.92 & 21.19 & 0.02 &  0.28 & 0.03 \\
   13 &  203.43 &  5.92 & 20.47 & 0.02 &  0.35 & 0.03 \\
   14 &  561.95 &  5.95 & 22.53 & 0.06 &  0.30 & 0.09 \\
   15 &  401.76 &  6.94 & 21.36 & 0.04 &  0.30 & 0.05 \\
   16 &  400.46 & 12.52 & 19.40 & 0.01 &  0.08 & 0.02 \\
   17 &  397.26 & 17.49 & 20.58 & 0.01 &  0.48 & 0.02 \\
   18 &   78.37 &  6.96 & 20.66 & 0.01 &  0.06 & 0.02 \\
   19 &  677.36 &  7.24 & 21.40 & 0.02 &  0.32 & 0.03 \\
   20 &  532.09 &  7.86 & 20.43 & 0.01 &  0.83 & 0.02 \\
\enddata
\end{deluxetable}

\section{The Color-Magnitude Diagram}

\subsection{Overall Morphology}

The analysis of the CMDs, presented in Figure \ref{fig2}a and 
Figure \ref{fig2}b for the CLUSTER and the FIELD sample respectively, shows
that the overall morphology of the main stellar populations in the NGC 1866 
region is well-defined. In particular, in Figure \ref{fig2}b, some 
typical features of the population mix in the field are evidenced 
\citep[see also the discussion by][]{wal95,ber92}:

(i) the field multi-population main sequence (MS) extending up to 
 $V\sim 16$ merges into the MS of NGC 1866

(ii) a pretty well-defined giant branch of a presumably old 
\citep[some Gyr, see][]{wal95} stellar population is clearly visible at 
 $(B-V) > 0.7$ and $V < 21$.

(iii) a red giant clump is located at $V \sim 19$, indicating the existence
of an intermediate-age population having $t > 500$ Myr, 
\citep[see][]{bro89,wal95}

On the other hand, from the comparison of Figure \ref{fig2}a and 
Figure \ref{fig2}b the
main features of the cluster population are also visible:

(iv) the bright He-burning clump between $V \sim 15.5$ and $V \sim 16.5$
 $(B-V)\sim 0.5$ and $B-V \sim 1.0$

(v) a fairly well-populated AGB which extends up to $V \sim 15$ and 
 $(B-V) \sim 1.5$. In addition, some stars lying in the portion of CMD,
that we assume populated by the field, belong to the cluster and are
late AGB stars detected and studied in the near-IR \citep{fro90}. We
shall discuss this point later on.

These features have no correspondence in the FIELD sample. On the basis of
these considerations, we can define three main regions in the CMD:

(1) the cluster Red Giants (CLUSTER RGs) for $(B-V) > 0.4$ and $V < 17$

(2) the field Red Giants (FIELD RGs) for $(B-V) > 0.5$ and $V < 21$

(3) the cluster Main Sequence (CLUSTER MS) for $(B-V) < 0.5$.

All these regions are separated by dashed lines in Figure \ref{fig3}.

However, in order to better enucleate the main features of the cluster 
population with respect to the field ones, we divided the CLUSTER sample in 5 
concentric annuli:

\begin{itemize}
\item annulus 1: $ r < 29\arcsec~(80pix) $; 
\item annulus 2: $ 29\arcsec< r < 45.4\arcsec~(80-125pix) $; 
\item annulus 3: $ 45.4\arcsec< r < 90.7\arcsec~(125-250pix) $; 
\item annulus 4: $ 90.7\arcsec < r < 130.7\arcsec~(250-360pix) $; 
\item annulus 5: $ 130.7\arcsec < r < 181\arcsec~(360-500pix) $; 
\end{itemize}

Figure \ref{fig4}(a-f) shows the radial CMDs for the five annuli defined above 
(panels a to e) compared with the FIELD sample (panel f). As we can see, 
although conspicuous errors affect the photometry in this crowded region, the 
innermost annulus (annulus 1) is clearly dominated by the cluster population:
no features of the field populations are present in this CMD, while both the 
CLUSTER RGs clump and the AGB are clearly visible.
On the other hand, the FIELD RGs regions become, as expected, more and more
populated at growing distances from the cluster center (annuli 2, 3, 4, 5),
and the main features of this population (the red clump and the old red giant
branch) progressively become more and more evident (compare panel (c),(d),(e),
with panel (f) in Figure \ref{fig4}). Another feature, which can be easily 
evaluated from Figure \ref{fig4} (panels (a), (b), (c)) is the termination 
point of the cluster MS which is fairly well defined and turns to be 
located at $V = 16.8 \pm 0.1$.

\subsection{Photometric errors}

In order to estimate photometric errors, we considered the internal errors from
DAOPHOT combined with the uncertainty of the calibration procedure and, when
possible, frame to frame scatter (i.e. for the shallow frames). We report in
Figure \ref{fig5}(a-l) the mean photometric errors in V and (B-V) for each of 
the five annuli defined above, averaged over 0.5 mag-wide bins. As expected, 
larger errors occur in the most internal crowded regions (annulus 1) and at 
fainter magnitudes.

\subsection{The completeness}

Since one of the main goals of this paper is to present meaningful LF and 
population ratios, the treatment of the completeness of the observed sample is
a crucial problem: for this reason, we shall discuss this point in detail.

The problem of estimating the completeness factor for a stellar population 
comes out dramatically when increasing the crowding conditions of the observed 
field. On the other hand, the crowding conditions depend, to a zero-th order, 
on the distance from the cluster center: in the second place, on the seeing 
because in bad seeing conditions, it ismore likely to blend close 
stellar images. The final effect of this phenomenon is the loss of fainter 
stars and, eventually, the appearances of spurious bright objects, that could 
remarkably alter the resulting LF.

In order to (quantitatively) estimate the fraction of objects lost in each 
magnitude bin, $\lambda _{c}$, we carried out extensive \textit{artificial 
star tests}. This procedure has been described in detail in many papers 
\citep[e.g.][]{mat88,bol89,san96}. The basic idea is to add a number of 
``artificial stars'' (i.e. objects having the same characteristics of the 
\textit{real} stars in the frame) to the original frame, and to reduce once 
again the new \textit{enriched} frame. In the specific case of NGC~1866 
(as in the case of any LMC cluster), the fact that the cluster is almost 
completely contained within a single frame generates a large density gradient 
within the frame and requires that the completeness parameter be evaluated 
also as a 
function of the distance from the cluster center. For this reason, we adopted 
the sub-division in concentric annuli defined in Section 3.1 to quantify the 
dependence of the incompleteness on the distance from the cluster center 
(i.e. on the crowding conditions). For each annulus we generated a set of 
\textit{artificial} stars for each bin of magnitude (0.5 mag wide) over the 
entire magnitude range covered by our observations ($V \sim 15-24$). In order 
to avoid spurious crowding enhancement due to the enrichement procedure, only 
20-30 stars were added to the original frame in each iteration, and the 
procedure was then repeated many times.
A virtually identical reduction procedure was performed on the 
\textit{enriched} frames in order to ensure a homogeneous treatment of both 
artificial and original frames.
The output list of stars (reporting magnitude and positions) was then 
cross-checked with the list of artificial stars added to the frame. 
Each artificial star was considered \textit{``detected''} if the following 
criteria were satisfied: $ \Delta X < 2 pix $, $ \Delta Y < 2 pix $, 
 $ \Delta mag<0.3 $, where $ \Delta X $, $ \Delta Y $ and $ \Delta mag $ are 
the differences in position and magnitude of the recovered star with respect 
to the simulated one.

The completeness factor ($ \lambda _{c} $) in each annulus (where the crowding
density can be considered constant) and in each bin of magnitude was finally
derived by the ratio between the number of artificial stars 
recovered ($ N_{rec} $)
and the number of stars originally simulated ($ N_{sim} $): 
$ \lambda _{c}=N_{rec}/N_{sim} $.

The procedure was applied independently to the deepest images in both filters
and repeated many times: a total of more than $ 10,000 $ artificial stars
have been simulated. According to \cite{mat88}, the final completeness factor
($\Lambda$) to each point $ (V,B-V) $ of the CMD is 
$ \Lambda =\lambda _{c}^{V}\times \lambda _{c}^{B} $.
Following this procedure, the completeness level at the CLUSTER RGs region
is $\sim 100\%$ in both the filters, while the correction factor in
the five considered annuli for the CLUSTER MS are plotted in Figure \ref{fig6}.
The completeness of our sample is high ($ >80\% $) down to $ V \div B<22.5 $ 
in the most external regions (annuli 3,4,5), then start to significantly 
decrease in annulus 2 (being $ >60\% $ down to $ V\div B < 21 $), and drops 
rapidly down to $ 40\% $ in the central field (annulus 1), even at a 
relatively bright magnitude level ($ V \div B\sim 19 $).
For this reason, we decide to exclude the contribution of this very central
region (annulus 1) in deriving the LF.

\subsection{Field Subtraction}

The traditional zapping technique consists in gridding the two CMDs, 
``cluster+field'' and ``field'', and count the stars in each cell in the two 
diagrams after normalizing the areas and correcting for completeness. Then, 
an equivalent number of stars is removed from the single cells of the diagram 
of ``cluster+field'', on the basis of the number of field stars found in the 
diagram of the ``field'' alone. This technique has two major problems: 

\begin{itemize}
\item the \textit{a priori} gridding leads to the possibility that populous 
sequences are arbitrary cut by the cell boundaries, and one cell is more 
affected than others by the subtraction. In this way spurious sequences could 
be generated, as for example in the case of an almost vertical MS cut in half 
by the cell boundaries, with a slightly redder field population MS; 
\item if, within a single cell, the stars to be dragged out are chosen 
at random from the sample, clearly the probability of picking stars where the 
density is higher, i.e. in the vicinity of the sequence, is large. The 
result would be an artificial depletion of populous sequences.
A possible solution is to trace a mean ridge line, and attribute
weights to the points so that it is more likely to drag out stars lying
far from the sequence rather than close to it. 
\end{itemize}
In order to avoid these problems, in the following we adopted the method 
described by \cite{mig96}. This method considers, for each star in the 
``cluster+field'' CMD, its own cell, then counts the stars which fall in the
cell, in the two diagrams (``cluster+field'' and ``field'', respectively).

The size of the cells are determined on the basis of the error-box associated
to each star, i.e. 
{[}$\pm MAX(2\sigma _{V},0.200),\pm MAX(2\sigma _{B-V},0.100) ${]}.
A probability is then assigned to each star: 

$$p=1-MIN({{\alpha (N_{FIELD}+1)}\over {N_{CLUSTER}+1}},1.0)$$

where $\alpha$ is the ratio of the areas of the field and of the cluster,
 $N_{FIELD}$ and $N_{CLUSTER}$ are the number of stars corrected for
completeness in each cell of the ``field'' and in the ``cluster+field'' CMDs, 
respectively. Then, a random number $p'$ is extracted, between 0 and 1. 
If $ p'<p $, the star is taken, otherwise is discarded.

As already quoted in Section 2.1, in this paper we assumed as FIELD sample all
the observed stars lying at $ r>6' $ from the cluster center. This area 
measures $\sim 16.8 \arcmin\Box$.

\subsection{Comparison with published results}

As we said beforehand, in the literature there are at least three previous 
photometric surveys specifically aimed at studying the stellar population in 
NGC 1866: C89, B89 and the revision by BCP94. In these papers, the authors 
presented LFs based on independent photometric studies of the cluster: the 
region sampled in each of the three papers are all included in the field 
covered by our observations. Moreover, the CMD plotted in Figure \ref{fig2}  
and Figure \ref{fig3} is by far the deepest, most accurate and populous CMD 
ever published for NGC 1866 (compare our Figure \ref{fig3} with Figure 5 in 
B89 and Figure 7 in C89 or even Figure 2 in BCP94). For this reason we
will use our sample in order to establish the existence of any 
systematic difference, bias or incompleteness which may affect the 
previously published samples.

Magnitudes and positions of the stars for all the published samples have been
kindly provided by the authors in electronic form.

\subsubsection{Comparison with C89}

The field observed by C89 covers an area of $\sim 2' \times 3'$ roughly
centered on the cluster center (see Figure 1 by C89). Their pixel size is 
comparable to ours ($ 0.363 \arcsec/$pix). 

First of all, the sample observed by C89 has been reported in our coordinate
system. The cluster center assumed by C89 is located at
($ X_{C89}=507.82,Y_{C89}=515.42 $), slightly different from the one assumed 
in our sample. However, in order to be fully consistent
with the assumption of C89 we adopted, in the comparison, the above 
coordinates for the cluster center.

Then, we searched for any systematic difference in magnitudes and colors 
between the two samples over the whole region in common in our field and C89. 
The stars in common between the two catalogs have been used to derive the 
equations relating C89 magnitudes to our system. The equations are reported 
below: 

$$ V=V_{C89}+0.145+0.014(B-V)_{C89}$$
 
$$B-V=(B-V)_{C89}+0.014.$$

Once the magnitudes and the position of the original C89 list were
reported to our system, we performed an accurate search for the stars in common
in the specific area selected by C89 to construct the \textit{normalized}
LF. The field specifically considered by C89 in their analysis is 
located between $ r_{int}=85pix $ and $ r_{out}=160pix $.
This area ($ \sim 2.4  \arcmin\Box$) is represented as a horizontally 
striped region labeled 'C89' in Figure \ref{fig7}. Stars have been considered 
identified under the following assumptions 
$$\Delta X<2 pix, \Delta Y<2 pix, \Delta mag<0.3$$

Figure \ref{fig8} summarizes the results of the comparison: in particular, in 
Figure \ref{fig8} we plotted the CMD for the 652 stars found in common in the 
selected area (panel a), stars which have been found ONLY in our sample 
(panel b), and stars which have been measured only in the C89 sample (panel c).

According to C89, dashed lines delimit different regions of the CMD mainly
dominated by different populations (see their figure 7). In particular, 
\textit{(a)} the region dominated by the cluster Main Sequence (MS) at 
 $ (B-V)<0.5 $, \textit{(b)} the FIELD RGs at $ (B-V)>0.5 $ and $ V>17 $, 
and \textit{(c)} the CLUSTER RGs stars at $ (B-V)>0.5 $ and $ V<17 $. 
Moreover, the CLUSTER RGs region has been divided into three sub-groups 
(labeled A,B,C) as suggested by C89. It is worth noting that while C89, at 
first, suggested to use only giants counted in Group A and B 
(since \textit{``group C likely do not belong to the same population as those 
of group A and B''}),they later adopted (in normalizing the MS LF) always the 
whole CLUSTER RGs population (group A+B+C) (see their Table 1 and the further 
discussion in their Sect. 4).

The most relevant results of the comparison are:

(i) Only 4 stars (over 656 stars) have no counterpart in our catalog (two
of them are in the region of the CMD dominated by the field population -- see
Figure 8c)

(ii) The C89 list turns out to be seriously incomplete with respect to our 
sample, even at relatively bright magnitudes ($ V>18 $);

(iii) All the 39 bright RGs (BRG) stars found by C89 have been identified in 
our sample. In addition, we found another BRG (namely star 3023: $ V=16.5 $ 
and $ B-V=1.0 $) which has no counterpart in the C89 catalog 
(see Figure \ref{fig8}b).

(iv) A BRG (namely 1018 in the C89 catalog) looks like a blend of two stars:
one of the two blended components still lies, though at a fainter magnitude, in
the BRGs region (Sector B - See Figure \ref{fig8}b); therefore this fact does 
not affect the total number of BRGs.

\subsubsection{Comparison with B89}

B89 sampled 4 fields ($ 4'\times 2'.5 $ each) at different distances from
the cluster center. One of the fields was centered on the cluster
and was used by the authors to construct the LF. We checked the relative
photometric calibration with respect to B89, following the same procedure
described above (with the same assumptions). 

After applying the cross-correlation procedure, 1355 stars in common
have been found and have been used to obtain the following equations which
relate the B89 magnitudes to our system:

$$V=V_{B89}-0.05-0.12(B-V)_{B89}+0.11(B-V)^{2}_{B89}$$

$$B=B_{B89}+0.02-0.38(B-V)_{B89}+0.31(B-V)^{2}_{B89}$$

In this case, we adopted a second order polynomial to derive
the correction. This gives a smoother relation that takes into account
both the small corrections at bright magnitudes and bluer colors and the larger
correction at faint magnitudes and redder colors. However, the overall 
corrections are in general very small ($ \Delta mag < 0.1 $). The region 
selected by B89 for constructing the LF is an annular region ranging 
 $ 47'' < r < 134'' $ from the center. This area 
($ \sim 9.3 \arcmin\Box$) is indicated as a tilted, diagonal shaded
region, labeled as 'B89', in Figure \ref{fig7}.

Similarly to Figure \ref{fig8}, Figure \ref{fig9} reports the CMDs resulting from the 
comparison between our catalog and B89, in the region selected by B89 for 
constructing the LF.

In this case too, the main results of the comparison closely resemble the 
discussion reported in the previous section. In summary:

(i) All the 52 BRGs found by B89 have been identified in our sample, and as
previously found in the comparison with C89 an additional BRG has been 
identified in our sample (namely star id. 1018);

(ii) also in this case the B89 sample appears to be seriously incomplete for 
 $ V > 18.5 $.

Interestingly enough, B89 did not perform any statistical decontamination from
field stars in the region of the CLUSTER RGs and in normalizing the MS LF they
used all the 52 RGs identified.

\subsubsection{Comparison with BCP94}

BCP94 completed the B89 survey presenting a (V,V-R)-CMD for an extended region
around the cluster (see Figure 1 in BCP94). As in the previous comparisons, we
used their sample in order to find systematic difference in magnitudes and/or
colors. In this specific case, only the V magnitudes could be compared to our
photometry.The BCP94 V magnitudes match our system by applying only a 
small offset:

$$V=V_{BCP94}-0.05$$

A detailed analysis of the stars in common has been carried out only in the
region that the authors chose for constructing the LF. This area is the 
annulus limited by the largest dashed circle which complete the area surveyed 
by B89 and labeled as 'BCP94' in Figure \ref{fig7}. The extension of the 
region discussed by BCP94 is $ \sim 13.8 \arcmin\Box$. 
Figure \ref{fig10}
presents the CMDs obtained from this comparison. As can be seen, the limiting 
magnitude of the BCP94 sample is located at $ V \sim 19 $. It is interesting 
to note, however, that in this case, for $V < 19$, only few stars in the 
cluster MS region are present in our list and are missing from the BCP94 list. 
In particular, all the 65 BRGs found by BCP94 have been identified. 
Figure \ref{fig11} shows the comparison of the integrated stars counts 
between the 
sample presented here and the three previous samples
(C89, B89, BCP94, respectively). All stars lying in the sector of the CMD 
defined as MS in Figure \ref{fig3}, in each of the regions selected by the 
various authors have been used for this comparison. No corrections 
(neither for completeness nor for field contamination) have been applied to 
the star counts.

The ratio of the stars counted in each magnitude bin can be used to derive the
\textit{``relative''} completeness ratio of each sample with respect
to ours. The comparison shows that the completeness ratio decreases to less 
than  $ \sim 50\% $ at $ V\sim 19.3 $ for C89 and BCP94 and at $ V\sim 20 $
for B89, respectively.

\section{Results}

\subsection{The CLUSTER CMD: decontamining the observed sample}

The decontamination procedure described in Section 3.4 has been applied to
each of the 5 radial annuli defined in Section 3.1. The procedure has been 
applied only for magnitudes brighter than $ B \div V = 22 $, because the LF 
will be studied only down to the magnitude bin where the overall completeness 
drops below 50\%, to prevent problems due to statistical fluctuations close to 
the cutoff line. Moreover, field decontamination at faint magnitudes requires
too high completeness corrections, which could give rise to unreliable 
star counts.

Figure \ref{fig12} reports the CMDs in the radial annuli 1, 2, 3, 4 and 5 
before and after the field decontamination, respectively. As discussed in 
Section 3.3, the innermost region (annulus 1) will not be used in the 
construction of the LF and for this reason it will not be considered in the 
following discussion. Thus, the final adopted sample covers the region in 
the $FA$ field over a radial range $ 29''<r<180'' $ (annuli 2+3+4+5). 
This field has approximately an extension of $ \sim 37.6 \arcmin\Box$.

A particular attention has been given to the statistical decontamination of
the region dominated by the CLUSTER RGs, since this population is used to 
normalize the MS LF in the comparison with the previous works, and -this is
even more important- it is the test population that will be considered in the 
following model comparison. Seven stars in the FIELD sample have been found 
to lie in the
region of the CMD labeled as CLUSTER RGs, thus we expected $ \sim 16 $ field
stars lying in that area of the CMD. This leads to a  final number of 107 RGs 
after the statistical decontamination from the field stars that 
will be considered in the following discussion. Figure \ref{fig13} 
reports the final decontaminated CMD for the global adopted sample 
(obtained co-adding the CMDs of annuli 2, 3, 4, and 5). The dashed lines 
reported in Figure \ref{fig13} identify the regions defined in 
Figure \ref{fig3}. There are few stars beyond the red limit of the
CLUSTER RGs box, which are indicated as Late-AGB stars. These have been 
identified from the near-IR sample of \cite{fro90} as stars in
the thermally pulsing AGB phase. Since our models do not include the latest
stages of AGB evolution, and this evolutionary phase is very short within 
the lifetime of a star, we will not consider them in the giant sample in 
the following analysis and discussion.

\subsection{The MS stars Luminosity Function}

In order to obtain the Luminosity Function of the CLUSTER MS stars, we 
followed a standard procedure already used in other papers 
\citep[see, e.g.,][]{fer97}. First, we determined the mean ridge line (MRL)
of the MS was  from the statistically decontaminated sample plotted 
in Figure \ref{fig13}, by adopting an iterative procedure: the sample was 
divided into 
0.2 mag-wide bins, then, for each bin, the star distribution in $(B-V)$ 
color was computed, the mode of the distribution was assumed as first-guess 
equivalent color, and all the objects lying more than $\pm 7\sigma$ away from 
the mode were rejected. Secondly, the mode of distribution was re-computed 
until convergence. 
When constructing the LF, only stars lying, in each 0.2 mag-width bin, 
within $ \pm 7\sigma _{B-V} $ from the MRL were counted. However, since the 
sequence's width depends on the errors, we performed the star selection for 
the LF in each annulus defined in Section 3.1. 

The global ILF obtained by co-adding each ILF computed in each annulus 
(2 to 5) is presented in Figure \ref{fig14}. The two lines represent the
observed ILF before (dashed line) and after (solid line) the correction for
completeness, respectively. Errors have been computed accordingly to the 
following relation:

\begin{equation}
\sigma _{\textrm{bin}}=[{N^{1/2}\over \Phi }+{N\cdot \sigma _{\Phi }\over \Phi ^{2}}]
\end{equation}
 where $ N $ is the number of stars effectively observed in each bin, $ \Phi  $
the completeness factor, and $ \sigma _{\Phi } $ the associated error which
was determined from the r.m.s. of the repeated completeness trials (tipically 
less than 0.05).

In the same way, Figure \ref{fig15} reports the ILF normalized to the total 
number (107) of RGs in the statistically decontaminated sample (NILF). The 
global uncertainty of the NILF in each bin depends also on the statistics of 
the number of RGs used to normalize the ILF, accordingly to the relation:

\begin{equation}
\sigma _{\textrm{NILF}}=[{\sigma _{bin}^{ILF}\over N_{RGs}}+{N_{ILF}\cdot \sigma _{RGs}\over N_{RGs}^{2}}]
\end{equation}

\noindent where $ N_{RGs} $ is the number of RGs adopted for normalization,
$ \sigma _{bin}^{ILF} $ is the error in each bin of the ILF obtained from
eq. (1) and $ N_{ILF} $ is the number of MS stars in each bin of the ILF.

\subsubsection{Comparing the NILFs}

In Figure \ref{fig16} we compare the MS NILF, computed adopting our sample,
with the ones published in previous papers. We considered 
each of the region covered by previous papers on its own (see Figure 
\ref{fig7}). 
IN order to emphasize the number of RGs adopted in each region to 
normalize the MS ILF, we briefly summarize the above results:
the actual number of RGs found by our photometry in each area surveyed by 
previous photometries are, respectively: 40,53 and 65 in the C89, B89, and 
BCP94 area. Accordingly to the adopted field population and the relative 
area extensions, we expected 1,4, and 6 field stars, respectively, in the 
CLUSTER RGs sector of the CMD. This yields a final number of 39, 49 and 59 RGs,
which were used to normalize the MS ILF. 
It is particularly interesting to note that, apparently, both B89 and BCP94 
did not apply any correction in order to take into account the field star 
contamination to the number of RGs. This correction is 
particularly large (6 RGs) in the area covered by BCP94.

The original NILF for the three published samples have been reported (as dashed
lines) for comparison in Figure \ref{fig16}. They have been measured directly 
on the original published figures since none of the various authors reported 
the adopted LF in form of a list, and we could not reconstruct the 
completeness correction and the decontamination procedure applied by the
authors to the observed samples. However, beside differences in the limiting
magnitude, techniques and procedure, the NILF obtained in this paper
agrees, within the uncertainties, with those published by previous studies at
the bright end of the NILF ($V < 19$). Also, the fact that the decontamination
procedure has not been applied to the B89 and BCP94 original sample does not
seem to have affected the NILF, probably because, in both studies, the authors 
systematically over-estimate the completeness correction for MS stars.

Finally, we report the original NILF from C89 and BCP94 in Figure \ref{fig17}. 
The NILF obtained by B89 is not reported because it is fully consistent with 
the one from BCP94 by which it is superseded. We also draw, for comparison, 
the global NILF obtained for the whole adopted sample (plotted in Figure 
\ref{fig15}).
The NILFs agree with each other, thus suggesting that the procedure adopted
for the field decontamination and completeness correction does not introduce 
any substantial
bias in the derivation of the NILFs. There is a weak indication that the NILF 
obtained from the C89 area is slightly more populated, but this should be 
regarded as an effect of statistical fluctuations.
In fact, given the error bars, the three distribution are fully consistent 
within 1 $\sigma$, and if the normalization factor of C89 is enhanced from 39 
to 41, the relative LF matches almost perfectly the others. 
On the basis of Figure \ref{fig16} and Figure \ref{fig17}, 
we can conclude that the statement by BCP94 that the differences in the 
observed NILFs were due to statistical fluctuations in the choice of the 
samples seems confirmed.

\section{Comparison with theoretical models}

\subsection{Preliminars}

In order to compare the present data with theoretical models, some issues such
as metallicity, distance modulus, and metallicity must be discussed. 
In the next paragraph we 
report some of the most recent determinations, as well as our final choices.

\textit{Metallicity}: Most recent determinations of metallicity for NGC 1866
\citep{oli98,hil95} attribute to the cluster a value
$ [Fe/H]\sim -0.55\pm0 .30 $. It should be noted that \cite{hil95}
performed a relative calibration of NGC 1866 on NGC 330, so that, depending
on the value estimated for the latter cluster, the metallicity of NGC 1866 can
be enhanced to $ [Fe/H]\sim -0.35 $. 

\textit{Reddening}: Values of E(B-V) from the literature span from 0.05 
\citep{wal92} to 0.10 \citep{els91}. Most authors, anyway, 
adopt E(B-V) = 0.06 \citep[][,B89]{wal74,cas87}. 

\textit{Distance Modulus}: The distance of the LMC is subject of controversy
within the astronomical community. Even since the HIPPARCOS results, the 
problem is far from being solved once for all. The current values for 
the distance modulus span from 18.06 \citep{sta98} -which is, however, an
outlier-  to 18.70 \citep{fea97}.
Useful reviews of recent results on the distance modulus of
the LMS can be found in \cite{mad98}, \cite{fea99} and \cite{wal99}.

In the following discussion, we adopt 
 $Z=8\times10^{-3} ([Fe/H]\simeq -0.4)$, $Y=0.26$, and two visual distance 
moduli, namely $(m-M)_V=18.6$ and $(m-M)_V=18.8$, while the reddening will be
assumed to vary between 0.06 and 0.10 (which means that the true distance 
modulus will range at most between 18.3 and 18.6, having adopted 
$A_{V}=3.2E(B-V)$).

\subsection{The analysis}

We will focus our discussion on the LF because it provides a more 
robust tool than the distribution of the stars in the HR diagram 
(since it does not use the radii or, alternatively, 
the effective temperatures of the models, i.e. colors).
Since we want to minimize the influence of the corrections for crowding 
on the observed LF, we limit our analysis to the region brighter than 
 $M_V=2.6$. In fact, down to this magnitude, the global corrections 
(which have been included!) to each bin are always smaller than 10\%, 
with only annulus 2 being corrected for a factor 40\% in the faintest bin.
The global correction to the total is constrained below 8\% (see sect. 3.3
and 3.4).

The theoretical models have been computed with the evolutionary code FRANEC, 
whose earliest and latest versions are described, respectively,
in \cite{chs89} and \cite{csl98}, while the
latest input physics are described in \cite{scl97}.
Most models used here have been published by \cite{dom99}, while
others have been especially computed for this work.
The models computed with an artificially enlarged convective core (hereinafter
AECC models) have been obtained by increasing the size of the convective core 
by an amount equal to $0.25 H_{p}$. We refer to such models as AECC 
and not as models computed with overshooting, since the fit to this 
(or another) cluster can only give us information about the probable size of 
the convective core, not about the mechanism which determines this size,
as mentioned above.

All the theoretical luminosity functions have been obtained by means of 
a code which produces the synthetic clusters (O. Straniero et al., in
preparation): this is a sophisticated tool 
which can manage a) the photometric errors, b) the effect of the crowding, 
c) the presence of binary systems having an arbitrary mass ratio, 
d) an age spread, e) an arbitrary initial mass function and f) a large number 
of photometric bands.
Synthetic CMDS can be produced in the age range from few Myr up to several Gyr.
All the models of which our 
evolutionary database range from the main sequence up to the advanced 
stages of interest for this work, so that no artificial match between 
different evolutionary phases is necessary.

The various studies of the LF of NGC 1866 available in the literature make use
of the NILF, as discussed previously.
Such approach was certainly valid in the previous studies of this 
clusters since a solid and complete MS extending over several magnitudes was
not available. We have used the NILF in the previous sections in order 
to have a common ground with the previous papers on the subject.
However, since we now have a well-defined (and essentially complete) MS down 
to at least 3 mag below the TO, we have a chance to adopt the MS 
itself as  ``normalization factor'' instead of the He burning stars.
Our synthetic CMDs are populated until the observed total number 
of stars brighter than $M_V=2.6$ is reached and NOT until the total number of 
giants observed is reached. We believe this choice is, whenever possible, 
preferable to the other because: i)the total number of stars brighter 
than the quoted magnitude is largely independent on the assumed
size of the convective core, since it is dominated by MS stars whose
LF is only marginally affected by the size of the convective core; ii)
the number of MS stars is much larger than that of the giants and is 
therefore much less affected by stochastic fluctuations. 
It must be noted, however, that the total number of stars brighter than the 
quoted magnitude depends on the adopted distance modulus (since it influences
the limiting magnitude where we stop the counting), on the 
adopted present mass function (hereinafter PMF), and on the fraction of 
binary systems present in the cluster. We will discuss this below. The 
observed total number of stars brighter than $M_V=2.6$ 
in our sample, corrected for completeness, is 4004 for the
visual distance modulus $(m-M)_V=18.6$ and 4362 if $(m-M)_V=18.8$ is adopted; 
the total number of giant stars is 107.

Before discussing the fit to NGC~1866, we think it important to  
discuss  the dependence of the theoretical LF on various parameters. 
Panels a) and b) in Figure \ref{fig18} show, respectively, the 
differential and the 
integrated main sequence LFs for three ages, 80 (solid), 120 (dotted)
and 160 (dashed) Myr. Clearly, the TO
luminosity declines as the age increases while the overall shape of the LF
around the TO remains substantially unaltered within the considered age range.
 Panels c) to h) in the same figure present the same LFs shown in the 
first two panels, but with the inclusion of the giants: from these 
panels emerges that the clump luminosity also dimmers as the age increases.
It should be noted that the distribution of the stars within the clump shows
a double peak which flattens as the age increases. This behavior is easily 
comprehensible if we consider that the stars spend most of their He burning 
lifetime either at the red dip minimum and at the blue nose
of the blue loop, not in the middle. Since the older the isochrone the 
smaller the blue loop, it is also clear that the double peak feature 
disappears as the age increases.
Another clear feature in Figure \ref{fig18} is the 
magnitude gap between the TO and the red giant clump. This gap reduces as 
the age increases, because the clump luminosity dimmers faster than the TO
\citep[see][]{ccs90}. Figure \ref{fig19} shows, for the same three ages, the 
effect produced by an increase of a factor of two in metallicity on the 
differential LF: the left and right panels show, respectively, the MS and the 
total LF, while the solid and dotted lines refer, respectively, to the larger 
and the reference metallicity. The end effect of an increase of 0.3 dex in 
metallicity is a simulation of an older age; in fact, 
the clump luminosity becomes fainter, and both the double peak feature
within the clump and the gap between the TO and the clump itself are reduced,
while the TO becomes more luminous; the latter feature implies that the larger 
metallicity requires an older age in order to have the same TO luminosity.

The influence of the size of the convective core on the LF can be readily seen
in Figure \ref{fig20} where the solid lines refer to the AECC case while the 
dotted ones refer, once again, to the reference case. The increase in mass of 
the convective core has a deep impact on the LF: the region around 
the TO is modified because the TO luminosity increases and the 
AECC LF is much steeper around the TO. Also, the clump is significantly 
affected by the AECC, since it appears to be brighter and
flatter. It is interesting to note how the synthetic diagrams
populate the respective isochrones. Figure \ref{fig21} shows in each panel a 
different isochrone superimposed onto its synthetic diagram obtained by 
choosing a number of stars equal to the sample observed in NGC~1866,
i.e. 4004 stars brighter than $M_V=2.6$. The upper and lower panels refer,
respectively, to the two ages t=100 and t=200 Myr, while the left and right 
panels refer, respectively, to the reference (standard) and AECC isochrones. 
The MS stops being populated before the 
formal "theoretical" TO; this is due to the fact that these clusters 
(and NGC 1866 is even one of the most populous ones!) are not rich enough
to significantly populate the isochrone around the TO.
The effect is only marginally present in the standard case, but it is 
dramatic when the AECC model is considered. As a consequence, the difference 
between the two $M_{V}^{TO}$ (standard and AECC) is largely reduced if  
evaluated on the synthetic diagrams (i.e. on the basis of the most brilliant 
MS star) rather than on the isochrones. Let us explain this point in 
detail: the region between the small MS hook (see Figure \ref{fig21}) and the 
"theoretical" TO is populated by stars which are in the so-called 
\textit{overall contraction} phase: during this phase, the stars shift from a 
structure controlled by a central burning to one controlled by a shell burning.
The evolutionary timescale of this phase is largely determined by 
the size of the He core mass (which directly depends on the size of 
the convective core), in the sense that it scales inversely with the size of 
the He core. Thus, the larger the He core mass (i.e. the larger the previous 
convective core), the faster the evolution between the MS hook and the TO
(and hence the lower the expected number of stars in this region): this 
explains  why synthetic CMDs obtained by assuming a larger convective core 
have the TO region of the isochrone underpopulated with respect to the 
standard case.

The last effect we want to show is connected to the likely presence of 
a significant fraction of binaries. It has been recently found with HST
observations \citep{els98} that NGC 1818, a cluster similar to NGC1866, 
but slightly younger, has a conspicuous population of binary systems estimated
of the order of 30\%.

In our analysis, the presence of binaries is taken into account as follows:
 1) the mass of the first star comes, as usual,
from a random extraction weighed with a PMF having the same slope $\alpha$
used in the extraction of the single stars, 2)
the mass of the companion is extracted by assuming a gaussian distribution 
(of the masses) with a peak mass equal to that of the first star. In this
particular case, we chose a rather narrow gaussian distribution 
(semidispersion equal to 0.1), in order to maximize the effects due to the 
binary population. Clearly, the effect on the luminosity of the compound 
system is maximized if the two stars have similar masses
because, otherwise, one of the two would dominate and the total luminosity 
would practically be that of the most luminous of the two.
By the way, such an occurrence clarifies why the crowding behaves in principle 
similarly to real binary systems but is, in practice, quantitatively less 
effective: in fact, once the fraction of ``binaries'' is fixed, if one assumes 
that they are all due to the crowding, the masses (i.e. the luminosities) 
involved in the formation of the ``binary system'' are not constrained by any 
rule to be similar, and hence, as a consequence the less luminous of the two 
is hidden by the other which, in turn, will remain essentially unmodified.

Figure \ref{fig22} shows in the left panel a 
comparison between the integrated LF of the main sequence in the reference 
case (thin line, hereafter MSILF), and the one obtained including a 30\% of 
binaries (thick line) for an age of 120 Myr and a mass function slope 
$\alpha=2.3$.
The inclusion of a significant fraction of binaries alters the LF by
flattening the MS slope and steepening the LF of the TO region. This means 
that the presence of the binaries masks the real slope of the PMF,
in the sense that it appears flatter than it really is. The right panel 
shows, for example, how the reference MSILF already shown in the
left panel (computed with $\alpha=2.3$) is well reproduced by the one 
computed with $\alpha=2.6$ and a 30\% of binaries. Such occurrence will 
have a significant impact on the fit to the cluster (see below).

With the aid of the theoretical scenario discussed above, we can now readily 
turn to the analysis of NGC 1866. In particular, we shall first show the fits 
with the standard and AECC models, and we shall then discuss how they 
change by including a 30\% of binaries.

\subsubsection{Classical analysis with standard models}

\noindent The comparison between our standard models and NGC 1866 is 
summarized in Figures \ref{fig23} to \ref{fig26} for the assumed 
distance modulus $(m-M)_V=18.6$. In Figure \ref{fig23} we report, for the 
three ages 100, 120 and 140 Myr and five values of the PMF slope 
($\alpha= 2.0, 2.1, 2.2, 2.3, 2.4$), the comparison between the theoretical 
MSILF (histogram) and the cluster data (black dots). Although this is a 
standard way of comparing the theoretical and the observed integrated 
luminosity functions, we think it difficult to 
judge the goodness of a specific fit. Hence we computed, for each case 
shown in Figure \ref{fig23}, the differences (observed-theoretical), 
in units of $\sigma = (N_{obs}-N_{theo})/(N_{theo})^{1/2}$. 
These are shown in Figure \ref{fig24}. Hereafter, we 
will refer to this quantity as $\Delta\sigma$. Since it will 
constitute the basis on which we shall determine both the age and the PMF 
slope, $\alpha$, we discuss it in brief:  as the total number of stars brighter
than $M_V=2.6$ is equal to the observed number, the first point 
to the right end of each panel always corresponds to $\Delta\sigma=0$ 
(i.e. theoretical and observed number are identical). Moving 
toward brighter magnitudes, the other values for $\Delta\sigma$ will not 
necessarily remain around the zero (good match) but their trend
with the magnitude will depend on the current adopted values for both 
 $\alpha$ and the age: the larger the range in magnitude over which
$\Delta\sigma$ remains close to zero, the better the fit. Now, let us consider 
the various rows in Figure \ref{fig24}: the three panels in each row have been 
obtained by assuming a constant PMF slope and the three adopted ages. As
a general trend, an age increase tends to globally lower the sequence of 
points. By looking along the columns at Figure \ref{fig24}, one can see the 
effect of a changing $\alpha$ at fixed age. In this case, the global trend of 
 $\Delta\sigma$ is toward lower values as $\alpha$ increases. If we focus
on the first column in Figure \ref{fig24}, we can see that, for an age of 100 
Myr, an increase in $\alpha$ from 2.0 to 2.2 improves the fit:
all the points move down toward zero and a good match is obtained in the 
magnitude range 1.5-2.6; a further increase of $\alpha$ worsens the fit 
because the points corresponding to the brightest part of the MS 
still tend to move down toward zero, whereas those which were previously 
close to zero move progressively below it. Since, even in the best case 
($\alpha=2.2$), only a very small part of the MS can be fitted, we conclude 
that, for this age, models cannot fit the data well enough.
The same general behavior is obtained for the other two ages too; 
in particular, for t=120 Myr the "best" fit is obtained for $\alpha$ of the 
order of 2.2-2.3, while for t=140 Myr the "best" match corresponds to
$\alpha \sim 2.1$. At variance with the 100 Myr case, however, there is a 
good fit all along the magnitude range from -0.5 to 2.6 for both these ages. 
By increasing the age above 140 Myr one obtains a situation
similar to the one discussed for 100 Myr: no good fit is obtained for any 
value of $\alpha$ over a reasonably extended magnitude interval of the MS. 
By adopting a visual distance modulus of 18.6, we conclude that it is 
possible to constrain the age of this cluster in the range 120-140 Myr 
with $\alpha$ in the range 2.1-2.3, depending on the age. It is worth noting
that, by means of this technique, we have been 
able to constrain the age without uing -so far- the TO 
luminosity: on the contrary, we have shown (and used) that the part of the MS 
down to 3-4 magnitudes below the TO depends significantly on the age; 
an age derived in this way is much more robust than the one obtained by a 
fit to the most luminous MS stars, because these stars are very few and hence
are subject to heavy stochastic fluctuations.
The further fit to the TO luminosity remains confined within the age range 
found above. The various panels in 
Figure \ref{fig24} show that, in no case, it is possible to obtain a good fit 
to the ILF near the TO region. Standard models predict a 
smooth decline of the ILF toward the brightest portion of the MS 
while the observed ILF shows a much steeper profile. Since the theoretical 
ILFs preserve such a shape around the TO over an extended age range, 
we will not obtain a good fit for any standard model in any -reasonably wide- 
age range. 

We can now extend our analysis to the post main sequence evolution, 
i.e. the core He burning phase.
In Figure \ref{fig25} and Figure \ref{fig26} we show, respectively, 
the differential and integrated LFs of the total sample. By looking at 
the panels 
corresponding to the cases showing a good fit to the MS (see above) we find 
that, while the average clump luminosity is 
reasonably well reproduced, there are a number of discrepancies which can 
hardly be reconciled: in particular, the models predict a number of clump 
stars which is almost twice the observed one and also a significant 
($\sim$~0.6 mag) luminosity gap between the TO and the He clump, which is not 
observed.
We therefore conclude that it is impossible to 
obtain a reasonable fit to NGC~1866, having assumed a visual distance modulus 
of 18.6, with the use of standard models.

The same analysis has been made for a visual distance modulus of 
$(m-M)_V=18.8$. The logical steps followed to perform 
the fit are the same adopted for the shorter distance modulus. In this case,
we obtained a good fit to the MS only for $t = 100~Myr$, $\alpha = 2.0$, 
and $t=120~ Myr$, $\alpha \sim 1.9$, while
there is no good fit for ages outside this range, whichever $\alpha$ is used.
We show the best fit for this case in Figure \ref{fig27}. 
The same discrepancies already evidenced for the shorter distance modulus are 
still present: the shape around the TO region is not reproduced, nor are
the total number of giants and the gap between the bright end of the MS
and the clump. The adoption of a longer visual distance modulus does not 
improve the fit to the cluster, which remains rather poor.

We have thus obtained, once again, the classical old problem first raised by 
\cite{bec83}: i.e. standard models predict much more He clump stars than 
observed, with the difference that the validating sample is now much larger.

\subsubsection{Classical analysis with AECC models}

The fit with AECC models is shown in Figure \ref{fig28} to
Figure \ref{fig31} for the distance modulus $(m-M)_V=18.6$. 
Now, a good fit to the MSILF is obtained in the age range 200-220 Myr and 
 $\alpha \sim 2.1 \div 2.2$, but only for $t\sim 200$ Myr and 
$\alpha \sim 2.2$ there is also a reasonable fit to the TO. An analysis of the
post MS evolution (Figure \ref{fig30} and Figure \ref{fig31}) shows that 
the number of giants predicted for this "best" case is in good agreement with 
the observed one. However, it must be noted that the clump LF is not well 
reproduced since it is predicted to be, on the whole, less luminous than
observed.

The increase of the visual distance modulus to $(m-M)_v=18.8$ leads to
a largely poorer fit.
In this case, the MS may be fitted in the age range 180-200 Myr and 
$\alpha$'s in the range 2.0-1.8, but in no case do we obtain a fair 
fit to the TO region. The best overall fit (shown in Figure \ref{fig32}) 
corresponds to $t \sim 180$ Myr and $\alpha \sim 2.0$.
The number of giants predicted agrees with the observed ones, while the 
synthetic LF remains systematically below the observed and
shows a well-defined gap between the clump and the brightest MS stars, which 
is not observed (panels c) and d) of Figure \ref{fig32}).

We conclude that the AECC models may fairly fit the observed LF only if a 
"short" distance modulus is adopted.

\subsubsection{Non classical analysis with standard models}

We now discuss the influence that a consistent fraction of 
binary systems (30\%) has upon the fit to a cluster like NGC~1866.
As mentioned above, this value has already been found in
NGC~1818, by means of HST observations, and hence it is a reasonable 
starting value.

A good fit to the MS is obtained for $t=120 \div 140$ Myr and 
$\alpha$ between 2.4 and 2.6 when a distance modulus $(m-M)_V=18.6$ is assumed.
Although for both ages the shape around the TO is 
relatively well reproduced, an age of 120 Myr and $\alpha=2.5$ 
give the best fit to the data (see panels a) and b) in Figure
\ref{fig33}). Note that, at variance with 
the case without binaries, we now have a good fit to the TO region too, because
the presence of the binaries tends to steepen the ILF toward the bright end 
of the MS (see above). A further analysis of the giant
stars is partly contradictory: while the predicted number of giants (148) is 
$\sim 4 \sigma$ larger than observed (107), the clump luminosity is reasonably
 well reproduced (see panels c) and d) in Figure \ref{fig33}).

By increasing  the distance modulus to 18.8, we obtain a really good fit
(shown in Figure \ref{fig34}) for an age $t=100~Myr$ and $\alpha = 2.4$. 
The whole MS sequence, including its bright end, is well reproduced.  
The post MS evolution is also fairly well reproduced (see panels c) and d) in 
Figure \ref{fig34}): the number of giants predicted is within 2$\sigma$ of the
observed number and, even more important, the luminosity distribution of 
the giants closely matches the observed ones.

In conclusion, the adoption of models computed with a standard 
extension of the convective core may lead to a good fit to NGC~1866 if we 
account for a non negligible population of binary systems 
$(\simeq30\%)$ and adopt the ``long'' visual distance modulus 
(i.e. $(m-M)_0=18.5\div18.6$). The TO mass predicted for this  
``overall globally best'' case is $4.5~M_{\odot}$, while the He clump mass 
is around $4.8~M_{\odot}$.

\subsubsection{Non classical analysis with AECC models}

For sake of completeness, we also consider the case of AECC models with 
a 30\% of binaries. A good fit to the entire MS and the TO is 
now obtained for an age range $t \simeq 220 \div 240~ Myr$ and 
 $\alpha \simeq 2.4 \div 2.6$, if $(m-M)_v=18.6$ is assumed (see panels a) and 
b) in Figure \ref{fig35}). The number of predicted giants 
closely matches the observed one. Nonetheless, this fit must be 
rejected because the luminosity of the clump stars is largely missed:
panels c) and d) in Figure \ref{fig35} clearly show that the sum of 
the two effects, i.e. larger convective cores and a 30\% of binaries, leads 
to predicted giants with a luminosity far too weak in comparison with the
observed ones. The adoption of the longer distance modulus, which is not shown,
does not improve at all the fit since, here too, the ``theoretical'' He
clump is largely underluminous with respect to the observed one.

We conclude that the simultaneous inclusion of a binary component, together 
with the use of AECC models, does 
not provide an acceptable fit to the observed LF of this cluster.

Before closing this section, we shall comment the so-called 
NIMSLF (i.e. Normalized Integrated Main Sequence Luminosity Function) for the 
cases obtained by including a binary component. Panels e) in Figures 
\ref{fig33} to \ref{fig35} show the comparison between the observed 
(black dots) and synthetic (histogram) NIMSLFs; the open dots correspond to a 
2$\sigma$ changing in the observed NIMSLF.
Since this kind of figure gives a pivotal role to the number of giants, 
the fit obtained with the AECC+binary models clearly leads to 
the best fit. We have, however, shown that a deeper analysis, i.e. one which 
addresses all the features available in the observed LF, leads
to the conclusion that this case must be rejected. On the contrary, the fit 
obtained by adopting the standard models plus the binary component, although 
it gives a worse fit to the observed NIMSLF (they agree within 2$\sigma$), 
leads to a reasonably good fit to this cluster.

\section{Summary and conclusions}

In this paper, we presented a new set of photometric data which largely 
increase the samples presented up to now. We have also shown  
systematically how the various theoretical models and/or scenarios can fit the 
observed data. As a first conclusion, we find that the discrepancy first 
evidenced by \cite{bec83}, i.e. that standard models predict an excess
of He clump stars, is still present even with the latest 
standard models. The easiest and most straightforward way to avoid 
the discrepancy is to artificially increase the size of the convective core.
Unfortunately, a side-effect connected to the use of AECC models is a fainter
clump luminosity. If the convective core is larger, 
an older age is needed in order to fit the bright end of the MS, but in 
this case, the isochrone displays a fainter He-burning clump.

Another method used to reconcile the theoretical predictions with the present
set of data is to take into account the possible presence of a  
fraction of binary stars. Already C89 temptatively tried to 
check the effect of the presence of a binary population on the fit to their 
NILF. They concluded that the inclusion of even  50\% of binaries did 
not significantly alter the fit and, hence, binaries were not relevant
for the understanding of NGC~1866. A look at fig. 18 
of C89 shows that the inclusion of binaries increases significantly the NILF,
although not enough to match the observations.
Nevertheless, from that figure it was possible to show that binaries
mimick (of course only from this point of view) an enlarged convective core:
the observed ratio $N_{MS}/N_{He}$ is increased.
This effect was also noted by B89, who clearly stated in their final 
discussion that the presence of the binaries could mimick the effect of the 
overshooting on the NILF, and lead to a closer agreement between (their) data 
and the standard models. 

Note, however, that while models computed with AECC produce
a fainter clump luminosity, models with binaries do not do so.
Binaries alter a synthetic diagram in two ways: they overpopulate the 
bright end of the MS (thus reducing the ratio $N_{MS}/N_{He}$) and require
a steeper ``intrinsic'' PMF slope. This implies that, with a fixed total 
number of stars to be extracted, less massive stars are favored with respect 
to those around the TO and beyond: hence the predicted number of giants is
reduced. 

By including a fraction of 30\% of binaries,  we have shown that standard 
models can nicely fit the data. Of course, one could question the real 
occurrence of binary systems in NGC~1866. We have, in fact, two alternative 
scenarios:

a) if there are essentially no binary systems in NGC~1866, only models 
computed by significantly increasing the size of the convective core
can fairly fit the present sample of data. In this case, however, we are still
faced with the fact that it is impossible to fit the luminosity function of 
the clump stars. Such discrepancy increases together with the distance modulus.
In this scenario, the age of NGC~1866 would be $t \simeq 200$ Myr
and the PMF slope $\alpha \simeq 2.2$;

b) if there is a conspicuous fraction of binary systems, standard models 
provide a good fit to the present data. In this case, however, the preferred
visual distance modulus is $(m-M)_V=18.8$, i.e. $(m-M)_0=18.5\div18.6$, 
depending on the adopted reddening.
In this scenario, the age is $t \simeq 100$ Myr and the intrinsic PMF 
slope, $\alpha$, of the order of 2.4.

The choice between these two alternative scenarios can be done either by 
solving the problem of the distance modulus, because for each given modulus 
only one of them is favored, and/or by improving our knowledge of the 
fraction of binary systems existing in NGC~1866.

In our opinion a real advance in the knowledge of this (and maybe other)
class of objects requires an observational effort for a real 
determination of the fraction of binaries present in the cluster, 
more than a further increase in the data sample 
(which is, however, always welcome). As final remark, we hope that the incoming
Cycle 8 HST observations will finally solve the enigma of NGC~1866.

\section{acknowledgements}

We gratefully acknowledge Enzo Brocato, Cesare Chiosi and Antonella Vallenari 
for giving us computer-ready tables of their published photometries and their
isochrones, and for helpful discussions. We warmly thank Giuliana Giobbi for
editorial help, and the anonymous referee for useful comments.

\newpage
\figcaption[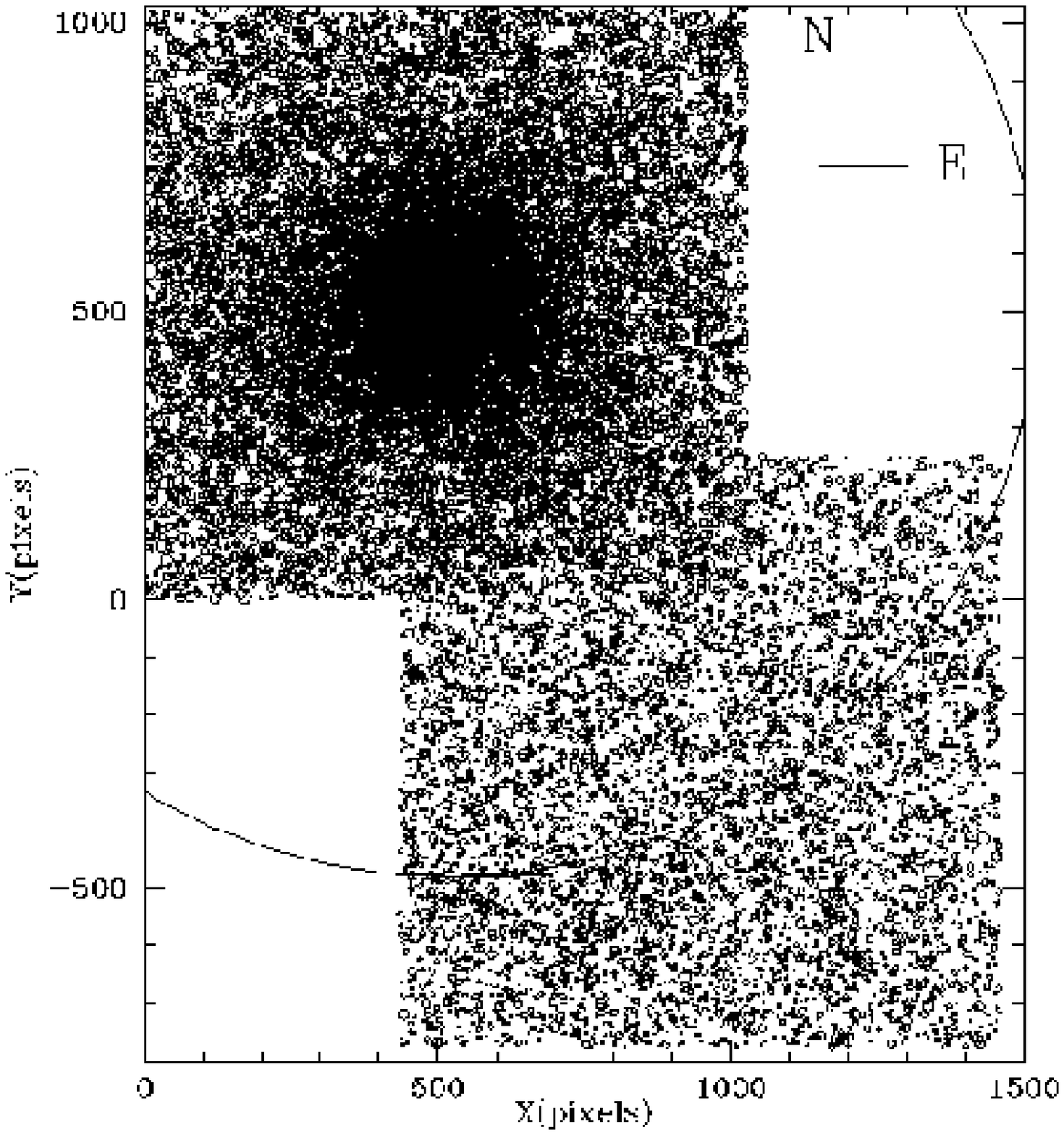]{Computer map of the two fields (FA and FB) observed
 in the region of NGC 1866. The coordinates are in pixels (0.363 arcsec/pixel).
 In this system the cluster center is located at pixel (520,520). The circle 
(at \protect$ r\sim 6'\protect $) delimits the FIELD sample (see text).
\label{fig1}}

\figcaption[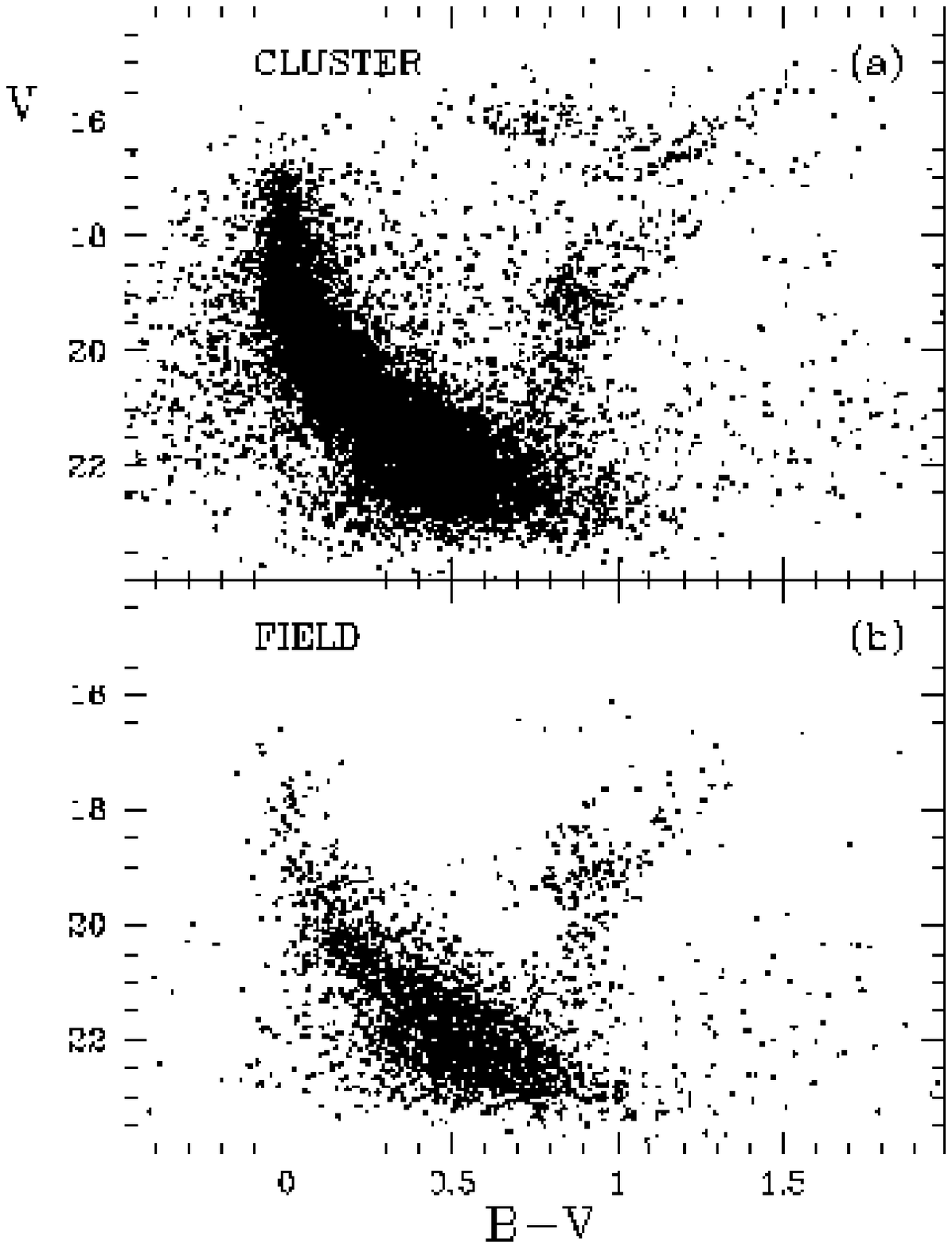]{CMDs of stars in: \textit{panel (a)} the CLUSTER 
sample (all stars in the field FA have been plotted); \textit{panel (b)} 
the FIELD sample (all stars in the field FB with \protect$  r>6'\protect $ 
from the cluster center are plotted.\label{fig2}}

\figcaption[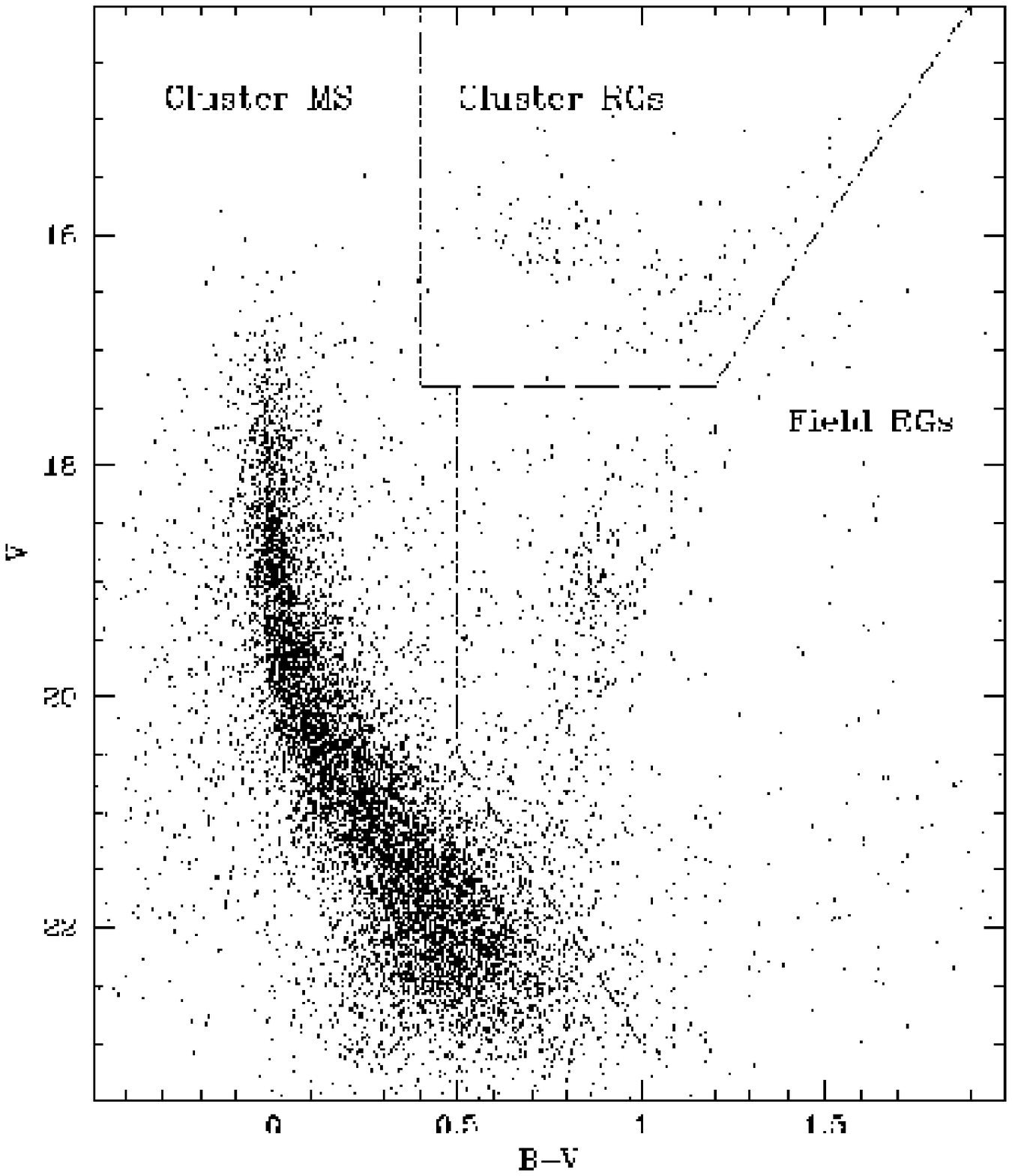]{CMD for stars in the CLUSTER sample. The dashed 
lines delimitate three regions in the CMD mainly populated by CLUSTER MS stars,
 CLUSTER RGs, and FIELD RGs, respectively.\label{fig3} }

\figcaption[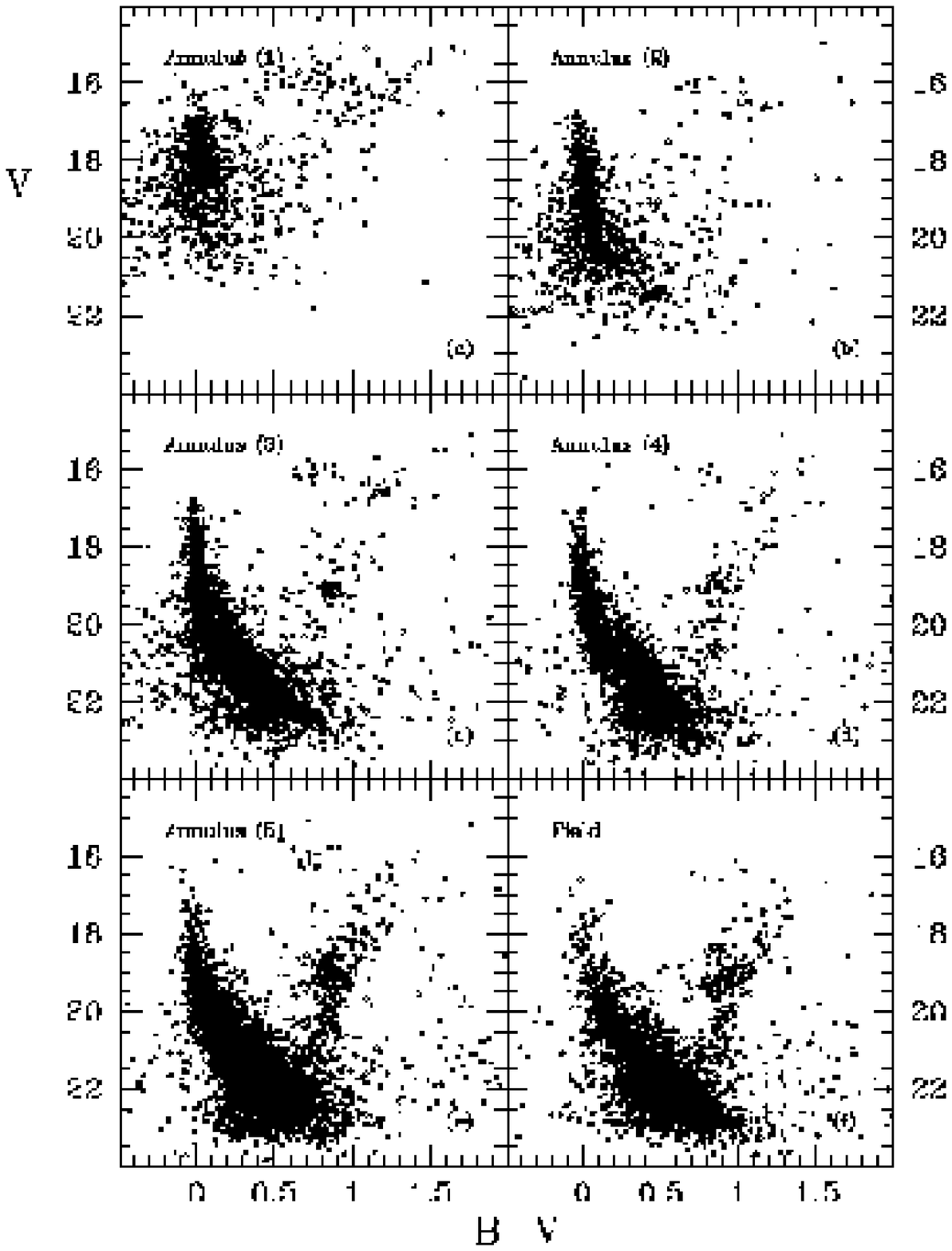]{CMDs for annuli at increasing distance from the 
cluster center (\textit{panels (a-e)}. The FIELD sample is plotted in 
\textit{panel (f)} for comparison.\label{fig4}}

\figcaption[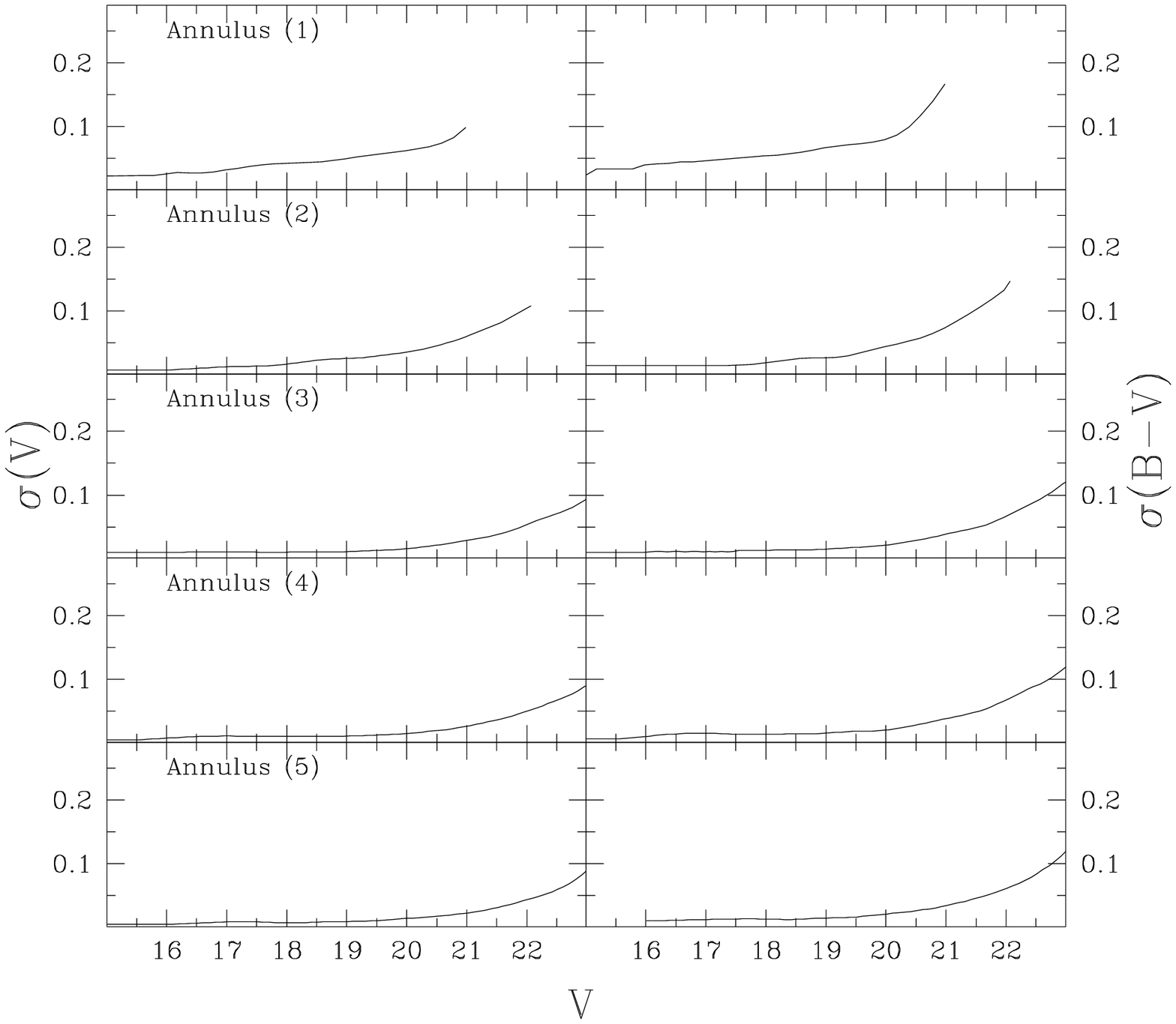]{Mean photometric errors in V magnitude and (B-V) 
color, for annuli at increasing distance from the cluster center.\label{fig5}}

\figcaption[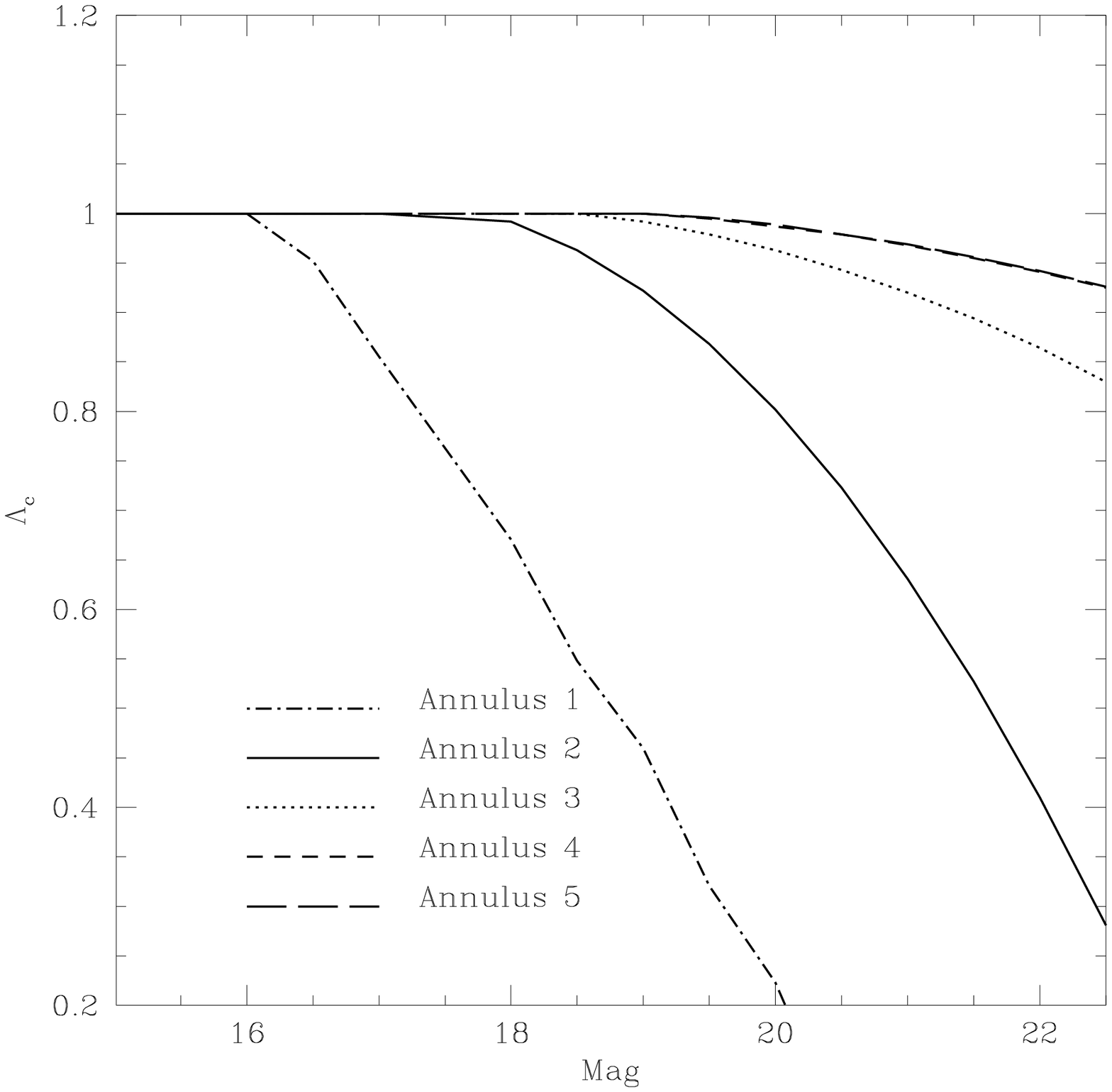]{Completeness curves for the MS stars in the 5 
annuli (1-5) at different distances from the cluster center as a function of 
the V magnitude as derived from \textit{artificial stars experiments}
\label{fig6}}

\figcaption[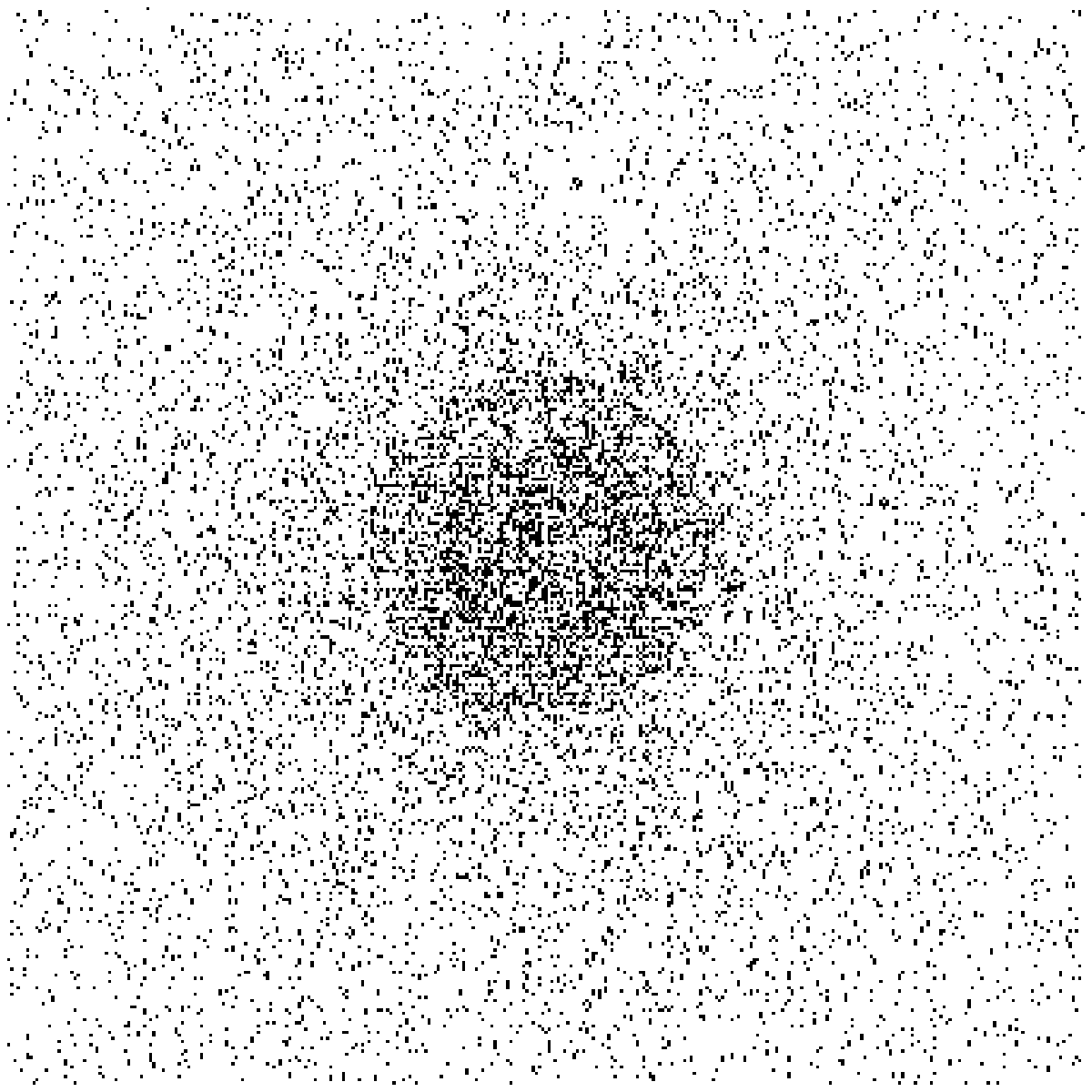]{The maps of the regions selected by previous 
studies (B89, C89 and BCP94) to construct the LF, are reported over the FA 
field of view.\label{fig7}}
\newpage
\figcaption[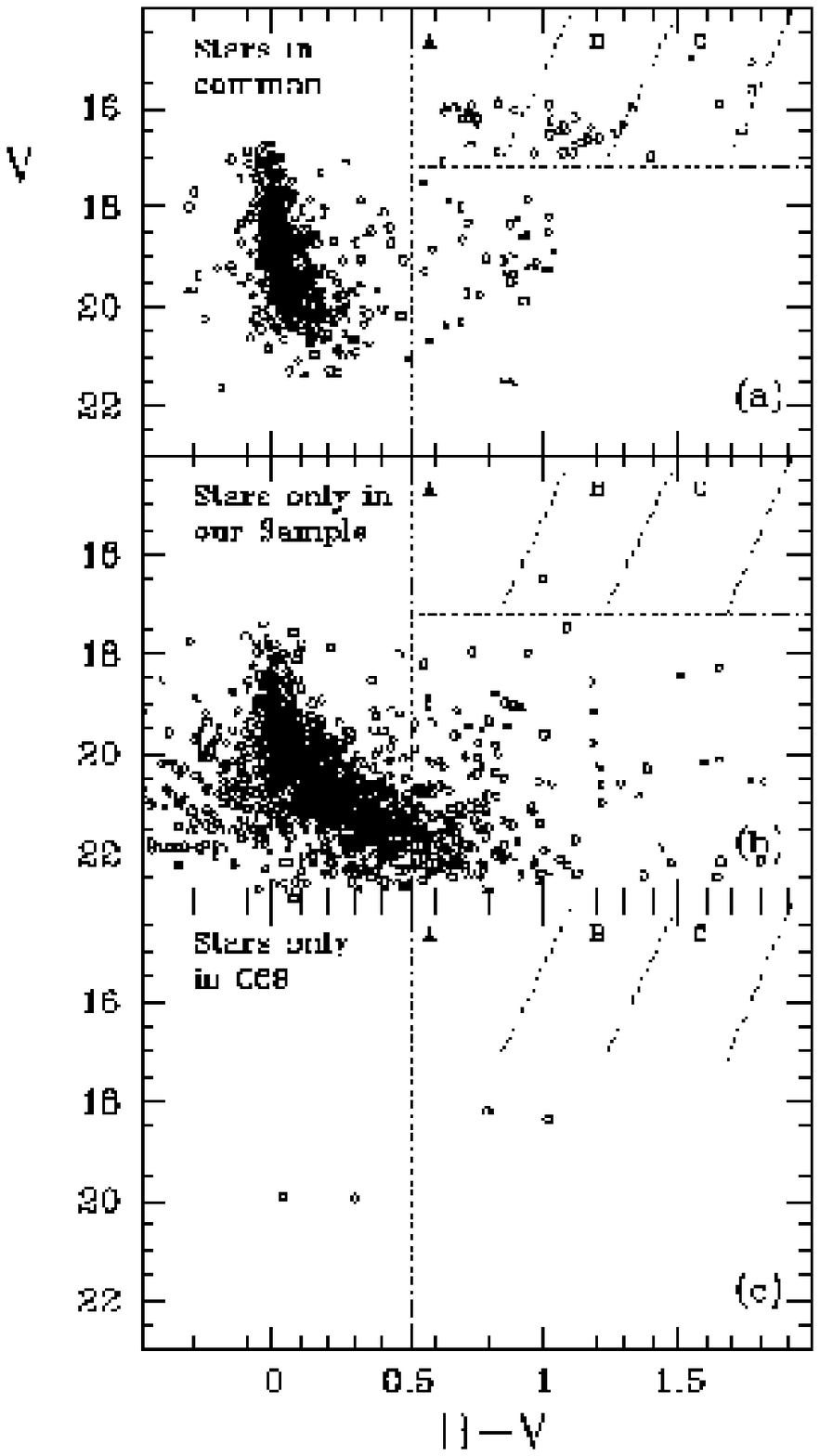]{Comparison with previous photometries: C89. 
\textit{panel (a)}: stars found in common; \textit{panel (b)}: stars found 
only in our sample; \textit{panel (c)}: stars detected only in C89. Only the 
region selected by C89 (see Figure 7) to construct their LF has been 
considered in the comparison.\label{fig8}}

\figcaption[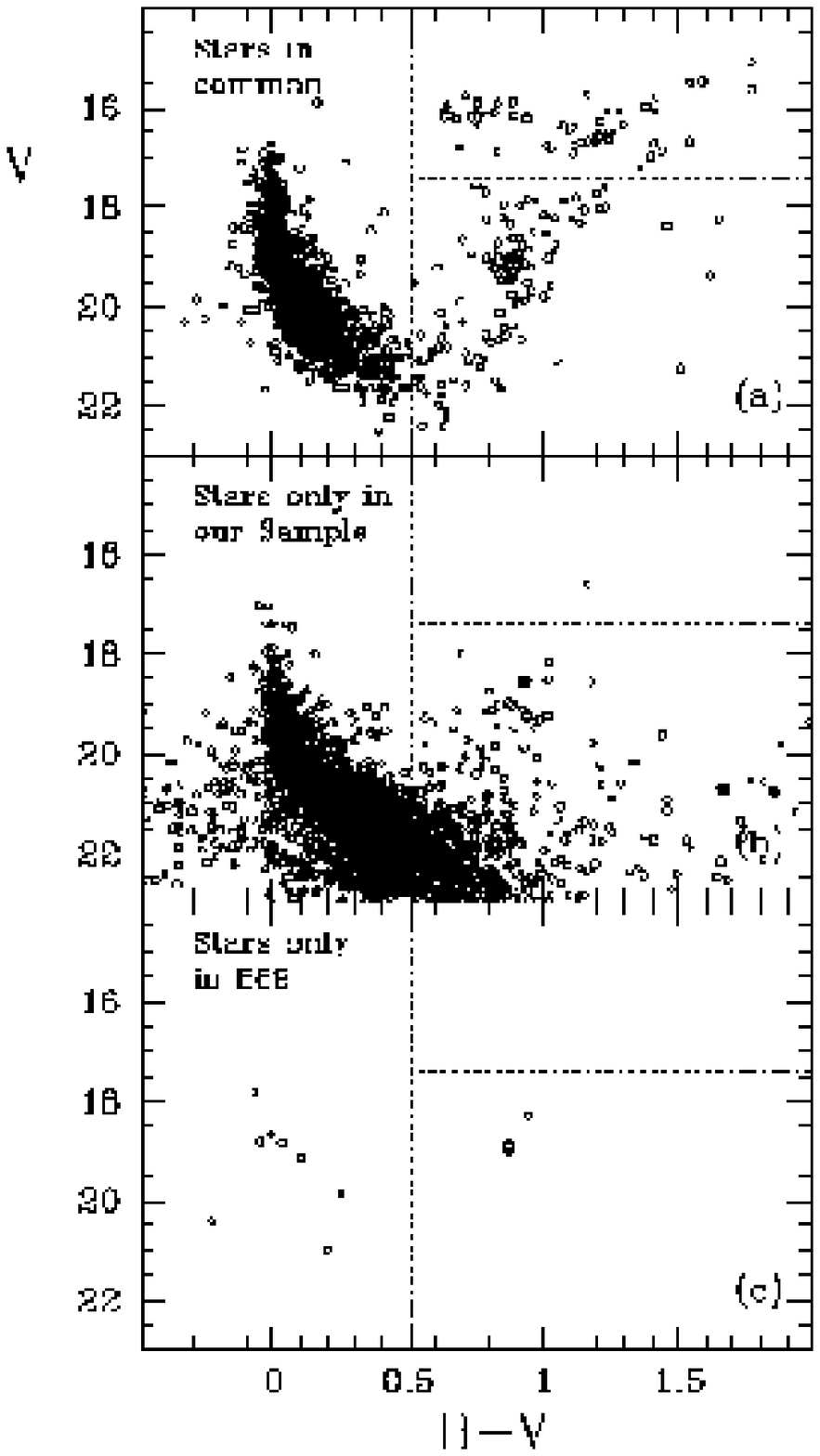]{Comparison with previous photometries: B89. 
\textit{panel (a)}: stars found in common; \textit{panel (b)}: stars found 
only in our sample; \textit{panel (c)}: stars detected only in B89. Only the 
region selected by B89 (see Figure 7) to construct their LF has been 
considered in the comparison.\label{fig9}}

\figcaption[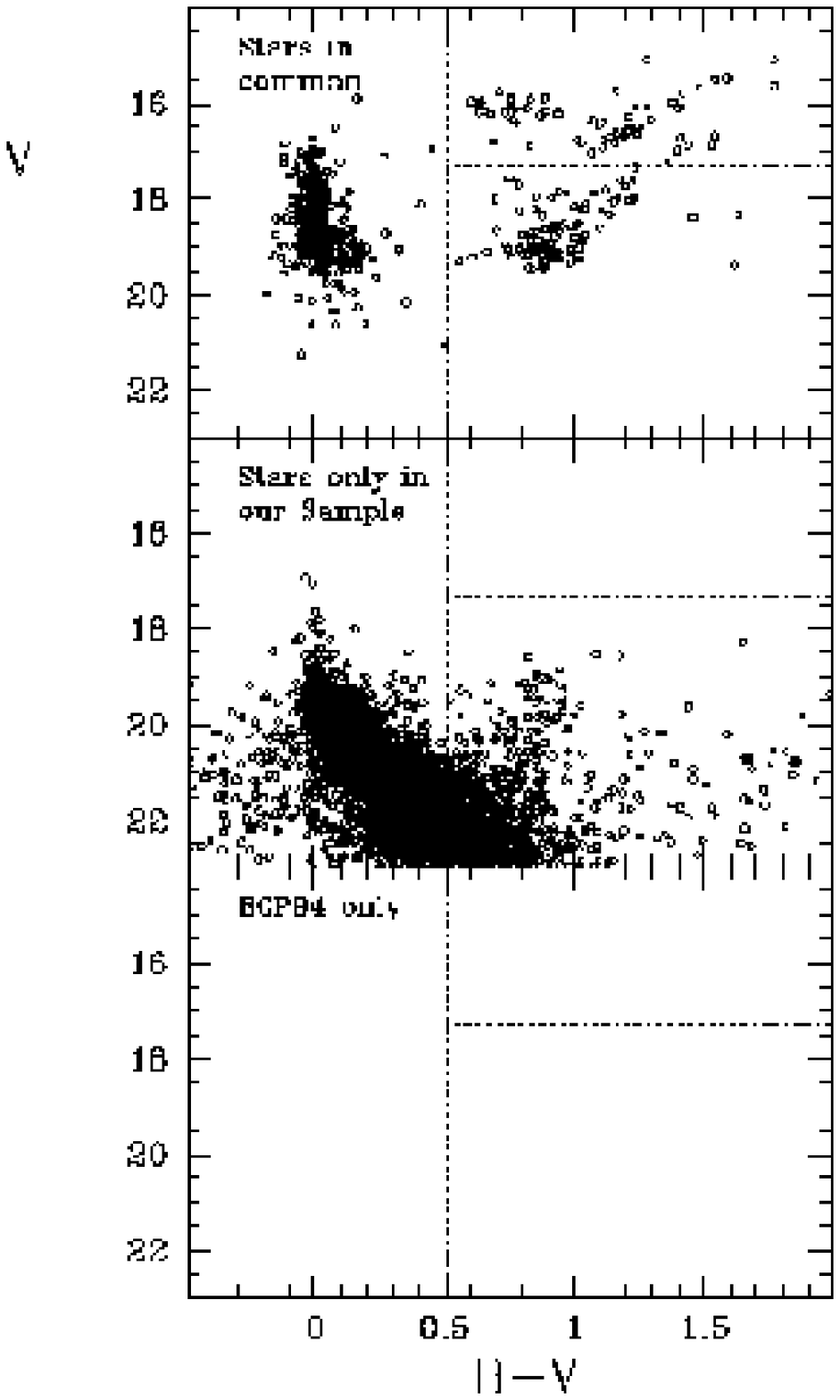]{Comparison with previous photometries: BCP94. 
\textit{panel (a)}: stars found in common; \textit{panel (b)}: stars found 
only in our sample; \textit{panel (c)}: stars detected only in BCP94. Only 
the region selected by BCP94 (see Figure 7) to construct their LF has been 
considered in the comparison.\label{fig10}}

\figcaption[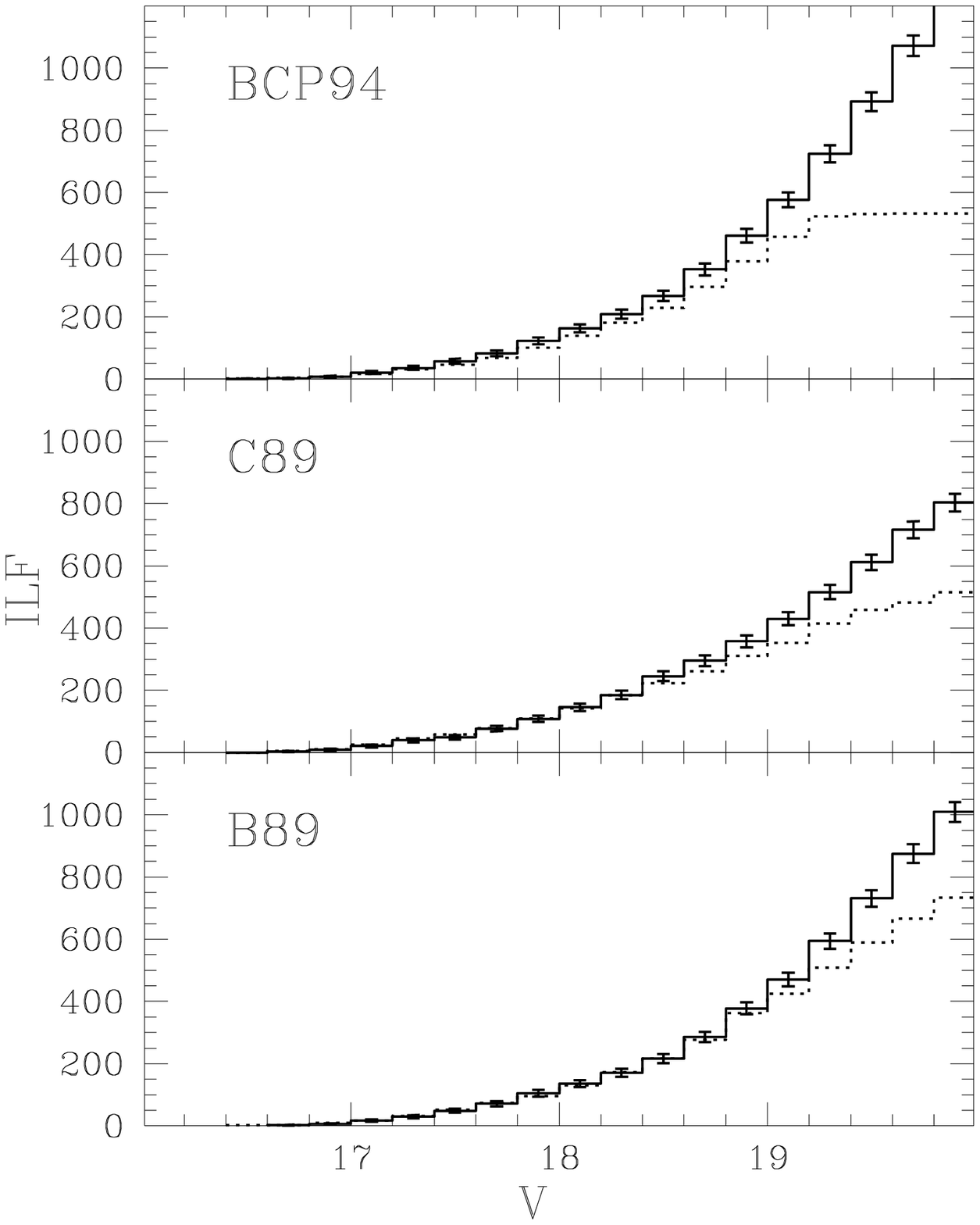]{Comparison between star counts obtained in the 
sample presented in this paper and the previous photometries (B89, C89, BCP94)
 \textit{panel (a),(b),(c)}, respectively. Only stars in the MS sector of the 
CMD have been counted. No corrections for completeness or field contamination 
have been applied to these counts. Only the regions selected by each authors 
(see maps in Figure 7) to construct their LF has been considered in the 
comparison.\label{fig11}}

\figcaption[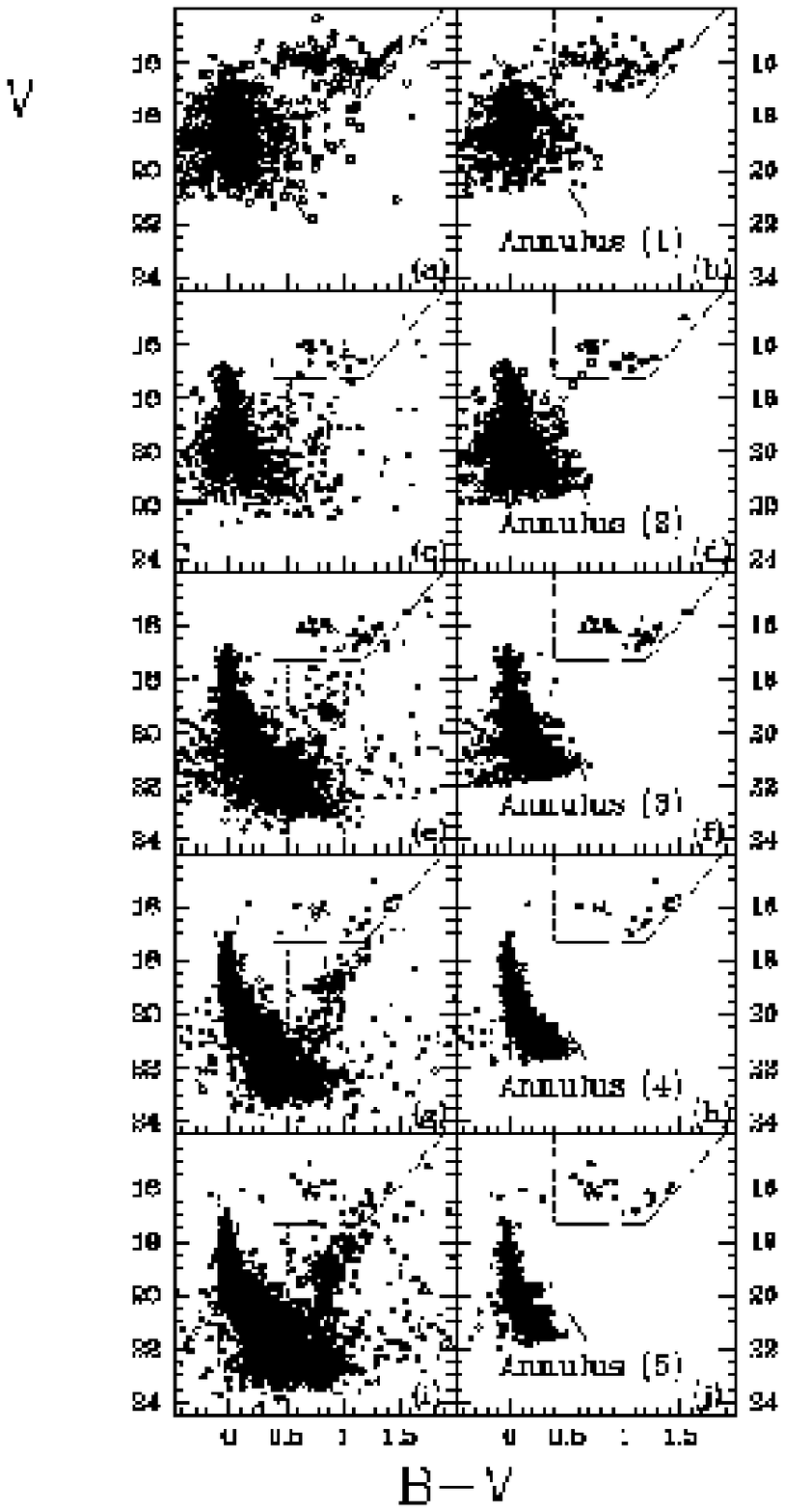]{CMDs for annuli at different distance from the 
cluster center (annuli (1-5)) before (left-hand panels) and after 
(right-hand panels) the statistical decontamination procedure.\label{fig12} }

\figcaption[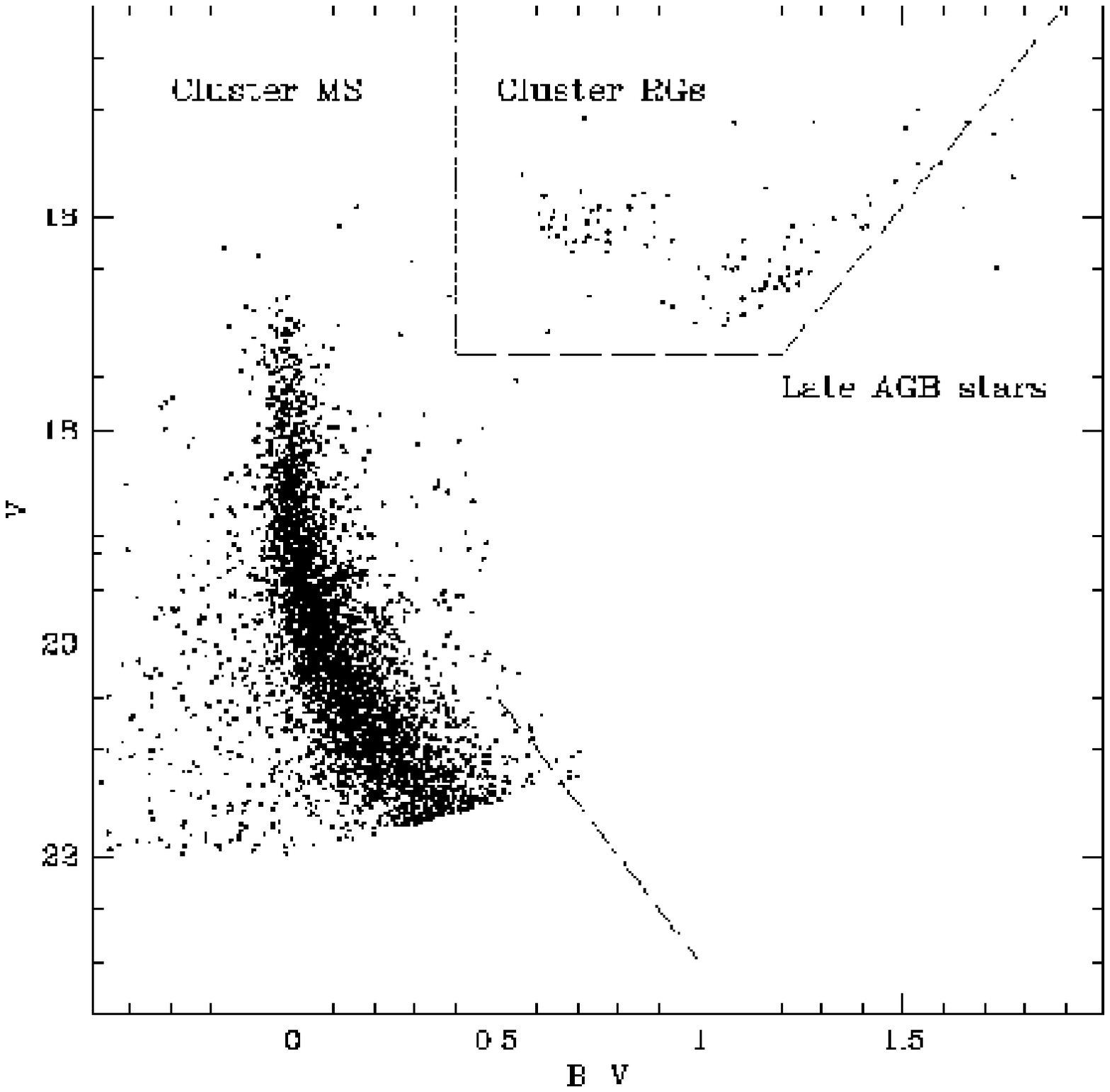]{CMDs for stars in the final adopted sample 
(annuli (2)-(5)) after the statistical decontamination.\label{fig13}}

\figcaption[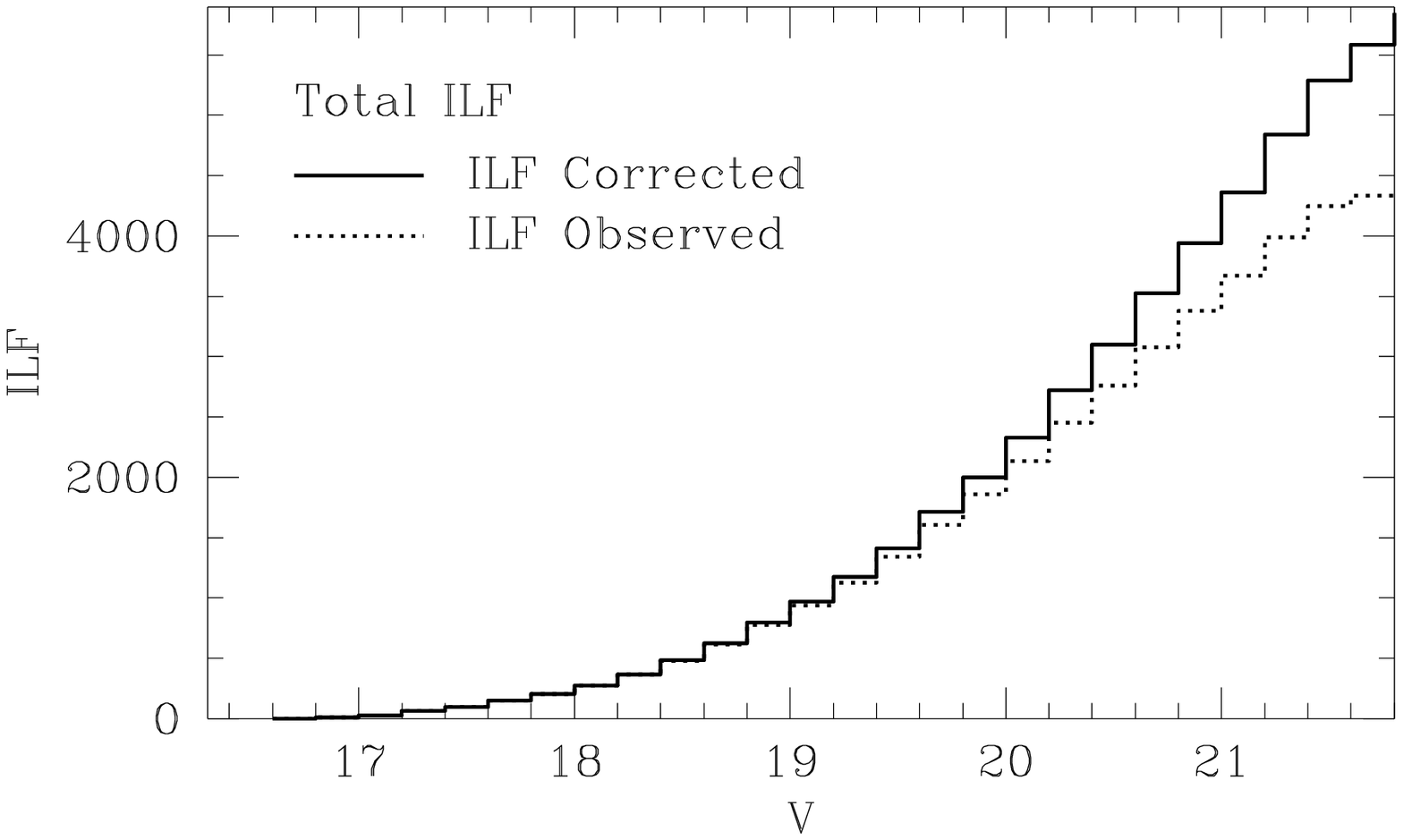]{ILF for MS stars with (heavy solid line) and 
without (dashed line) completeness correction\label{fig14}}
\newpage
\figcaption[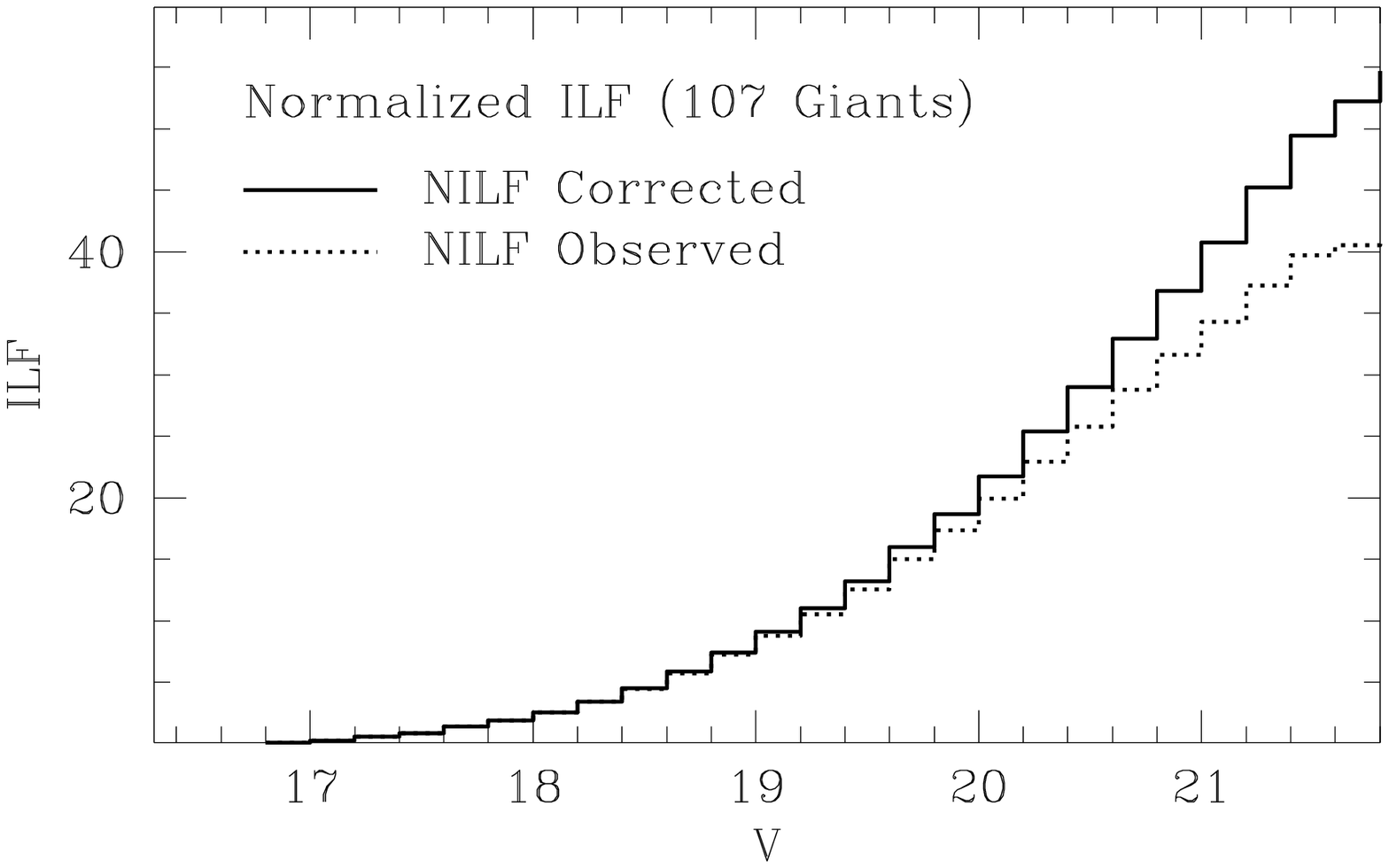]{ILF for MS stars normalized (NILF) to the total 
number of RGs. with (heavy solid line) and without (dashed line) completeness 
correction.\label{fig15}}

\figcaption[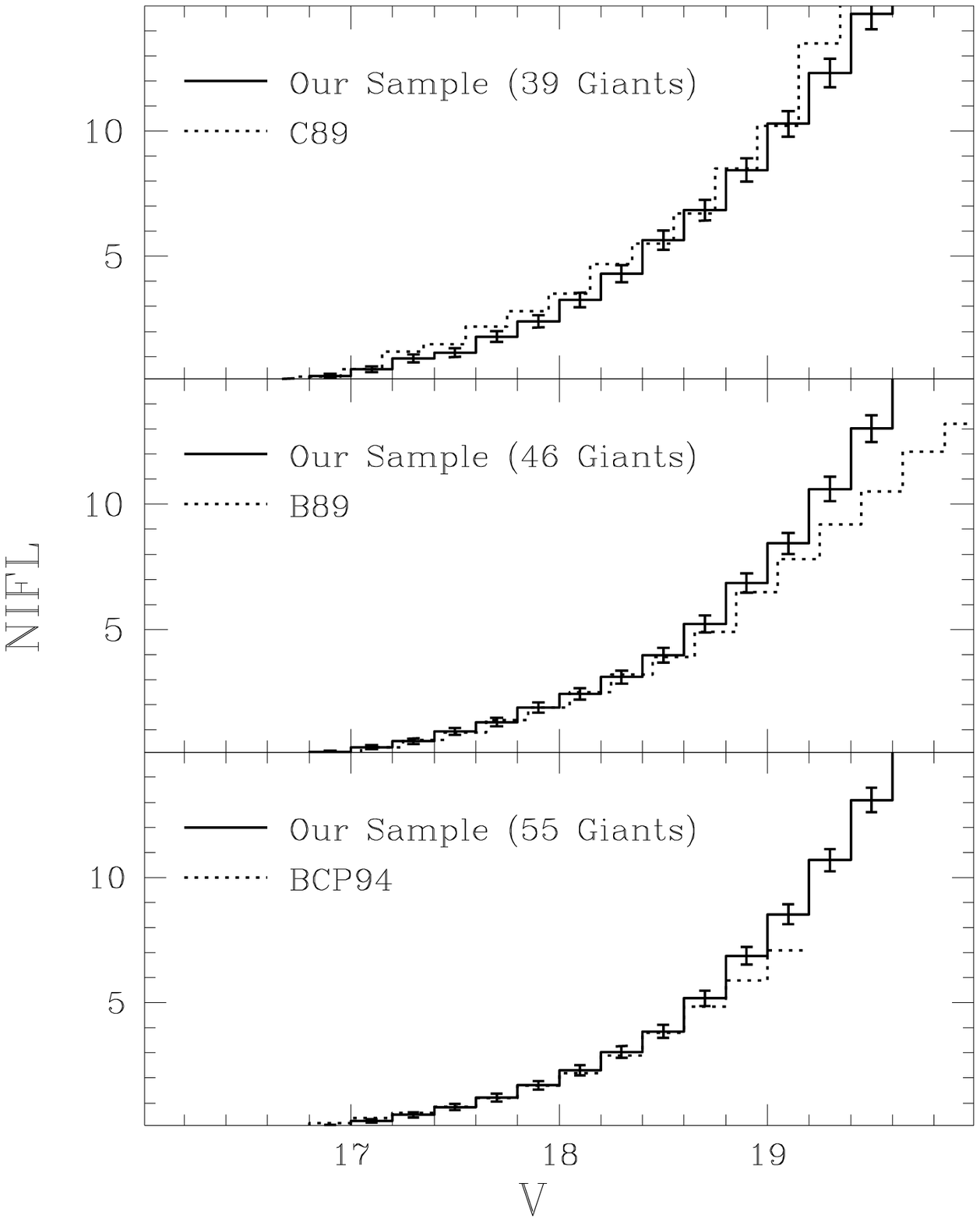]{Comparison with previous MS ILFs : C89, B89, 
BCP94, in panel (a),(b),(c) respectively. NILF based on the sample presented 
in this paper is plotted as heavy solid line. The number of RGs adopted in 
the normalization in area has been reported in each panel.\label{fig16}}

\figcaption[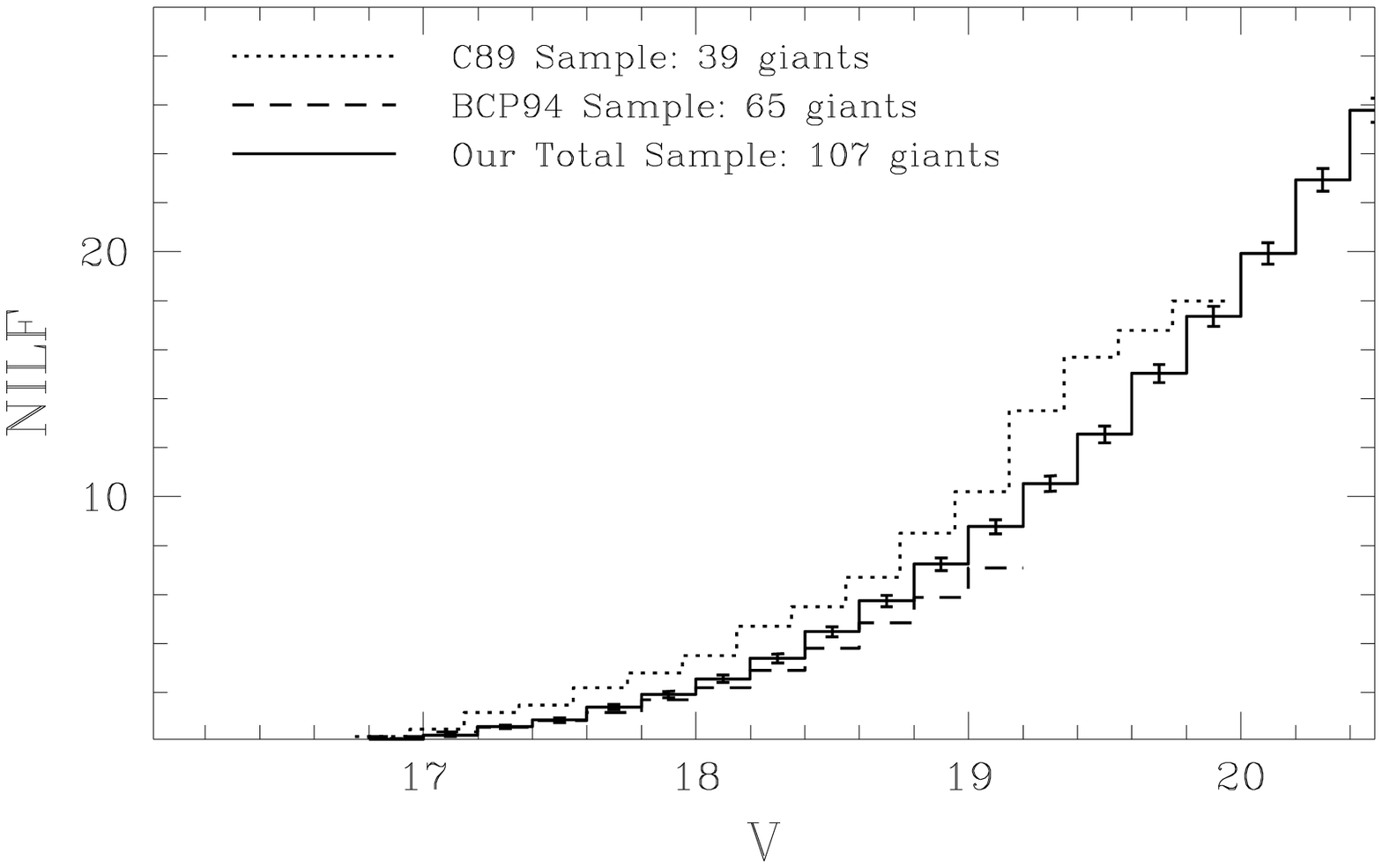]{Original MS-NILFs are plotted in the same panel 
to allow a direct comparison. Also plotted is (heavy solid line) the global 
NILF shown in Figure 15.\label{fig17}}

\figcaption[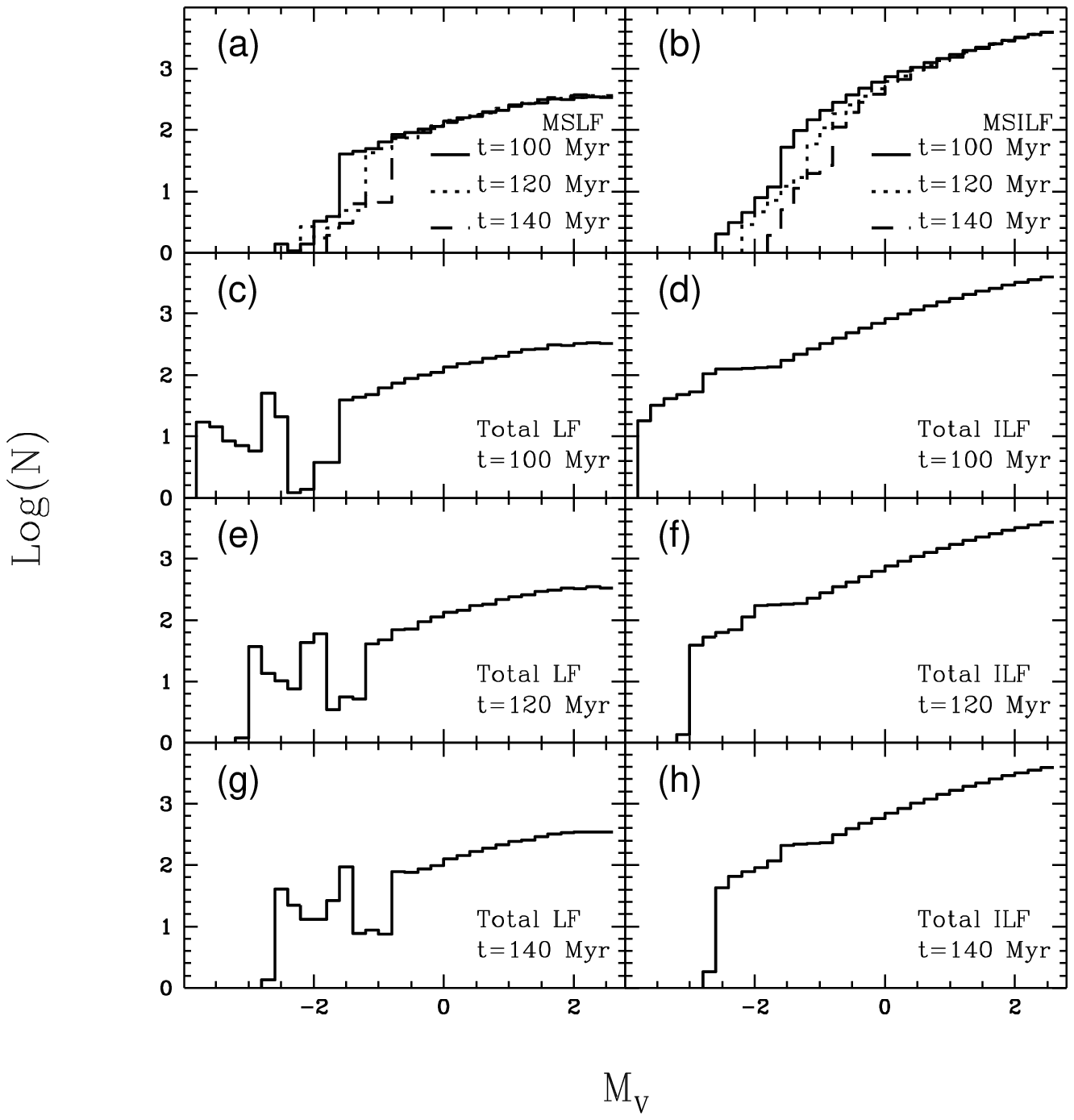]{Dependence of the model differential (left panel) 
and integrated (right panel) LFs on the age. Age used are t=100 Myr (solid),
t=120 Myr (dotted), t=140 Myr (dashed).\label{fig18}}

\figcaption[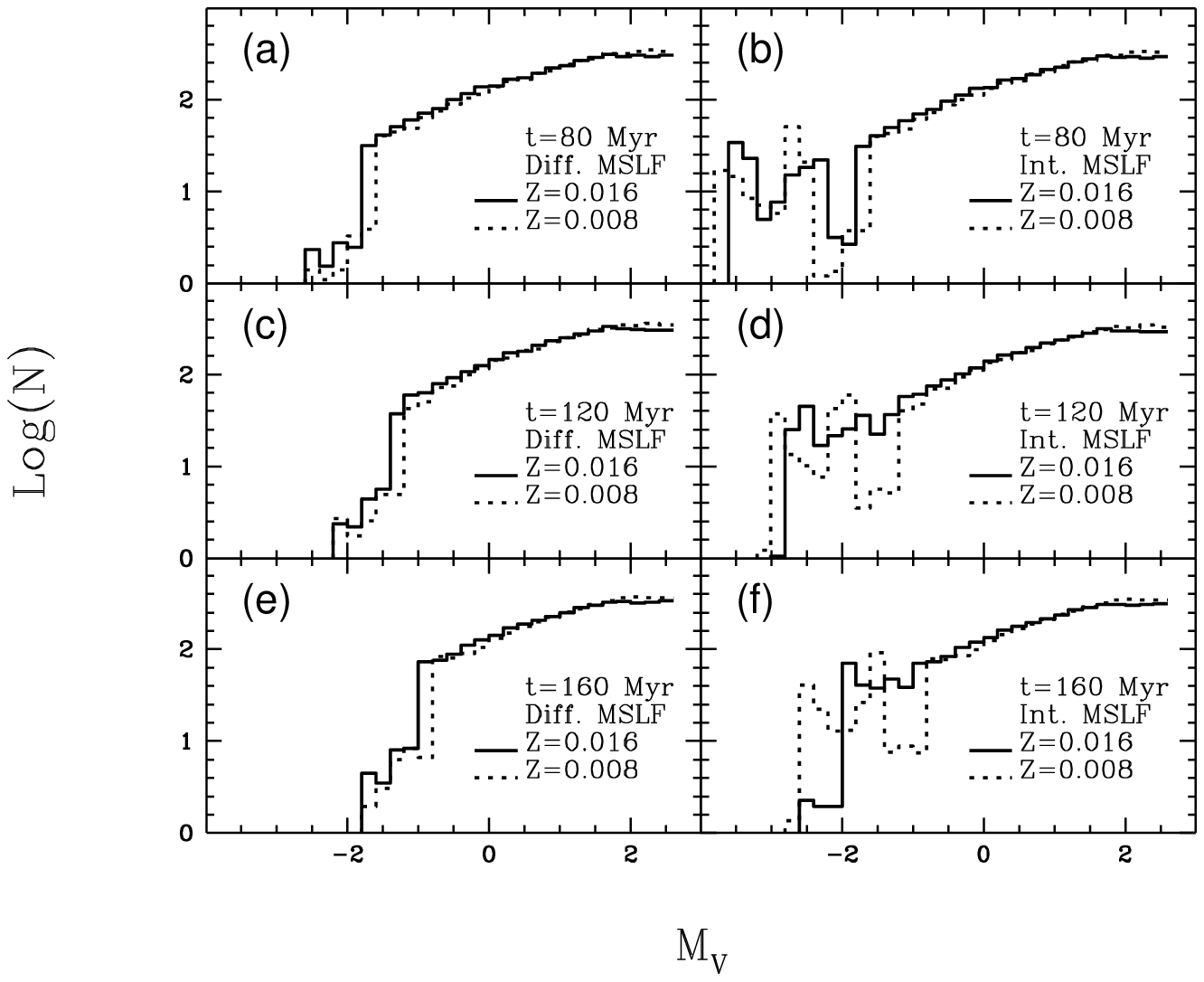]{Dependence of the differential (left panel)
and integrated (right panel) LFs on the metallicity. Lines refer to
Z=0.016 (solid) and Z=0.008 (dotted)\label{fig19}}

\figcaption[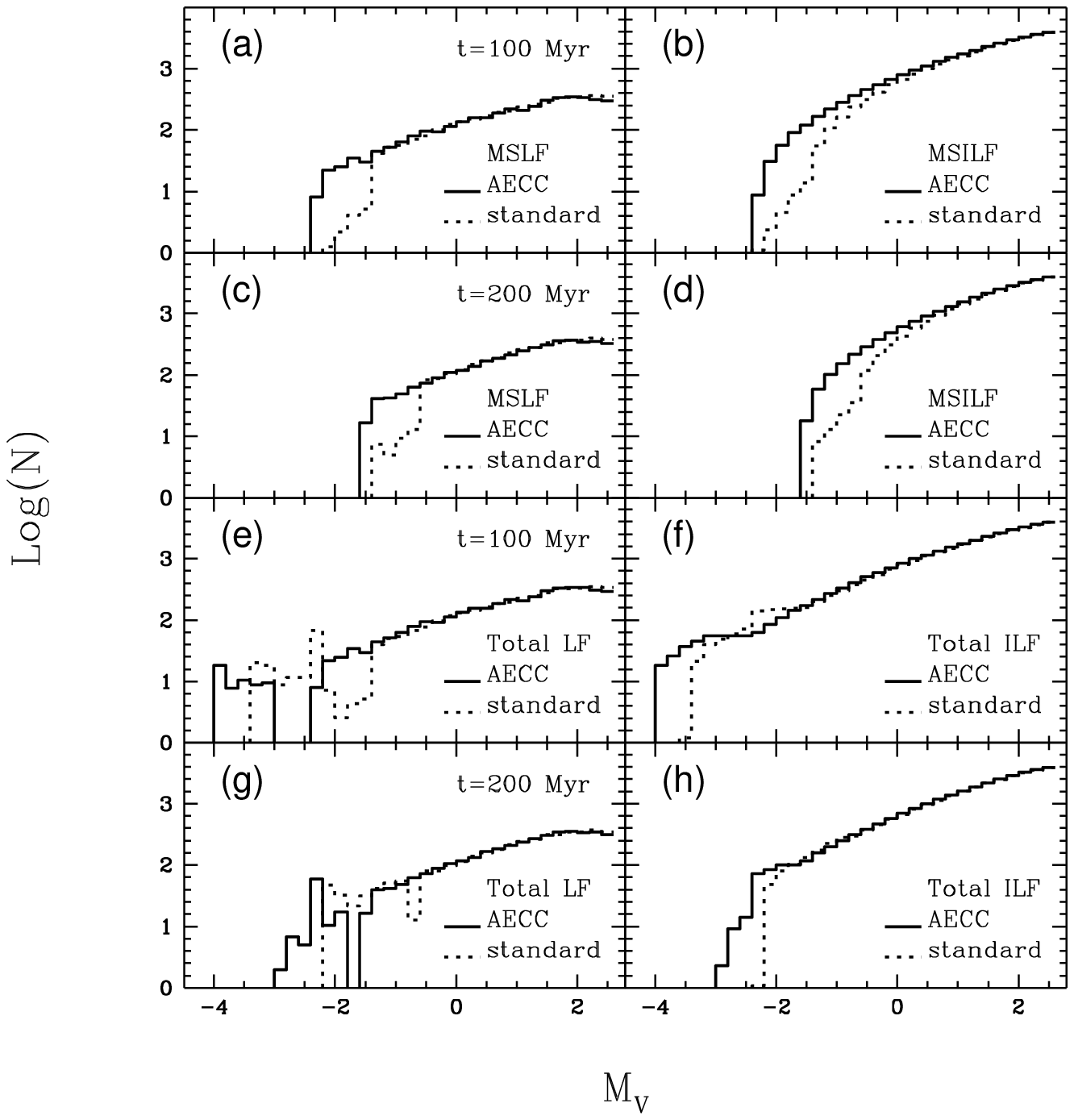]{Dependence of the differential (left panel)
and integrated (right panel) LFs on the size of the convective core 
(overshooting). Solid lines refer to AECC models, dotted lines to 
standard models.\label{fig20}}

\figcaption[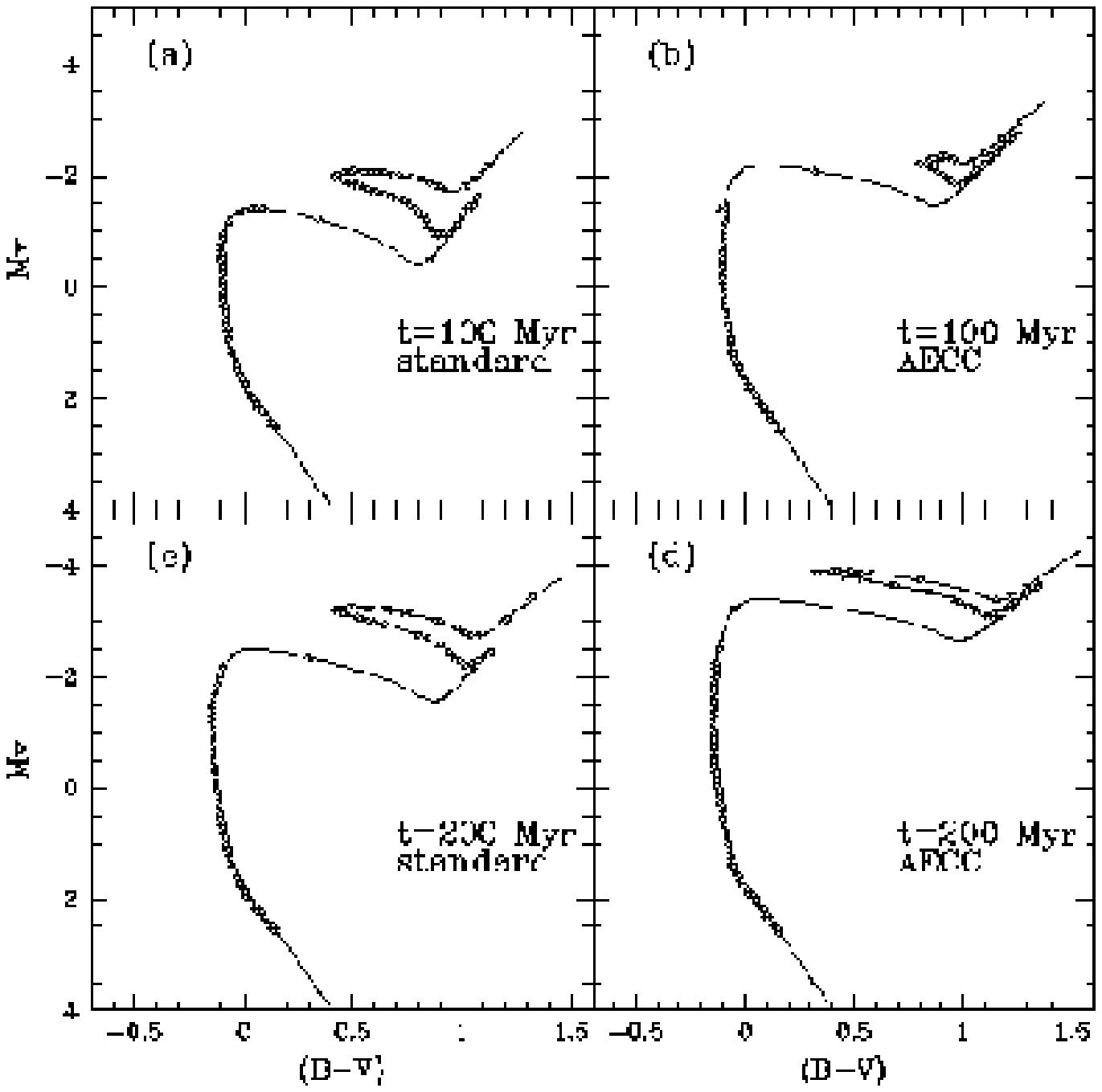]{Synthetic CMDs populating their respective 
isochrones for models with standard core size (left column) and models 
with AECC (right column) and for two different ages.\label{fig21}}

\newpage

\figcaption[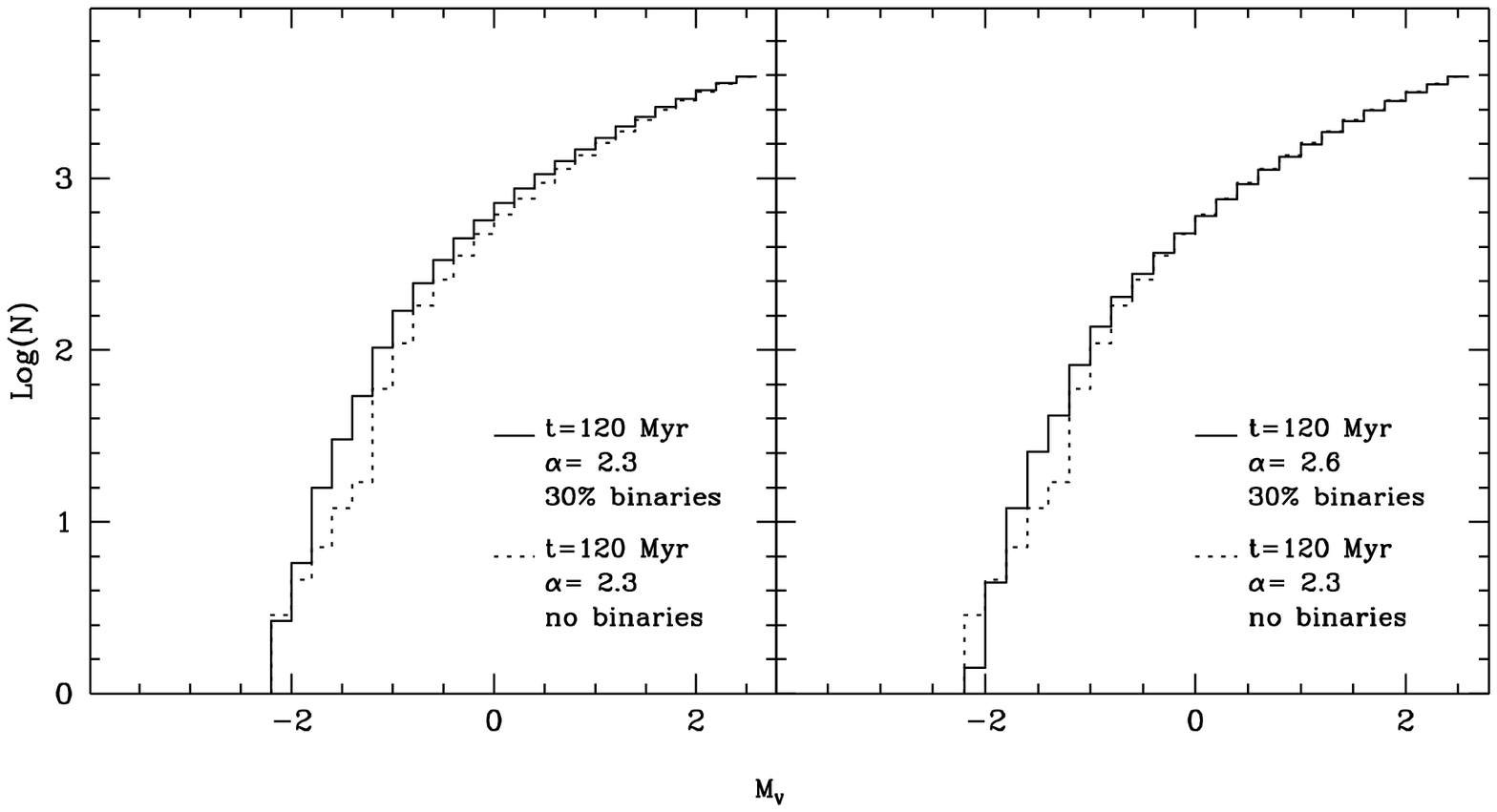]{Left panel: model MSILF with $\alpha = 2.3$ 
without binaries (dashed line) and with 30\% of binaries (solid line).
Right panel: the same standard case as in the left panel (dashed) is
reproduced with 30\% binaries and $\alpha = 2.6$ (solid).\label{fig22}}

\figcaption[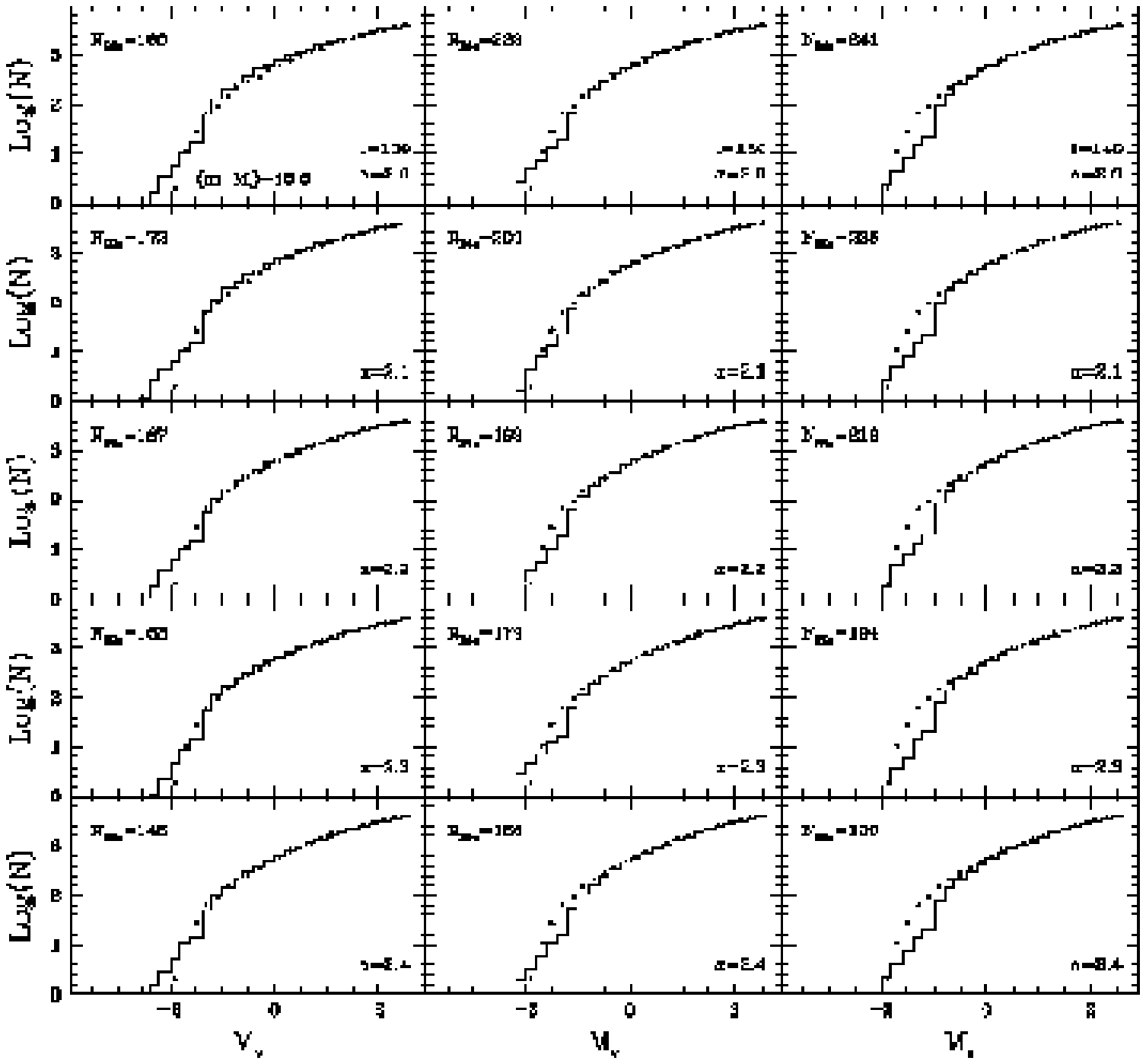]{Comparison between the theoretical MSILF 
(histogram) and the observed data (dots) in the standard case for three ages
(columns) and five values of $\alpha$ (rows) and $(m-M)_V=18.6$.\label{fig23}}

\figcaption[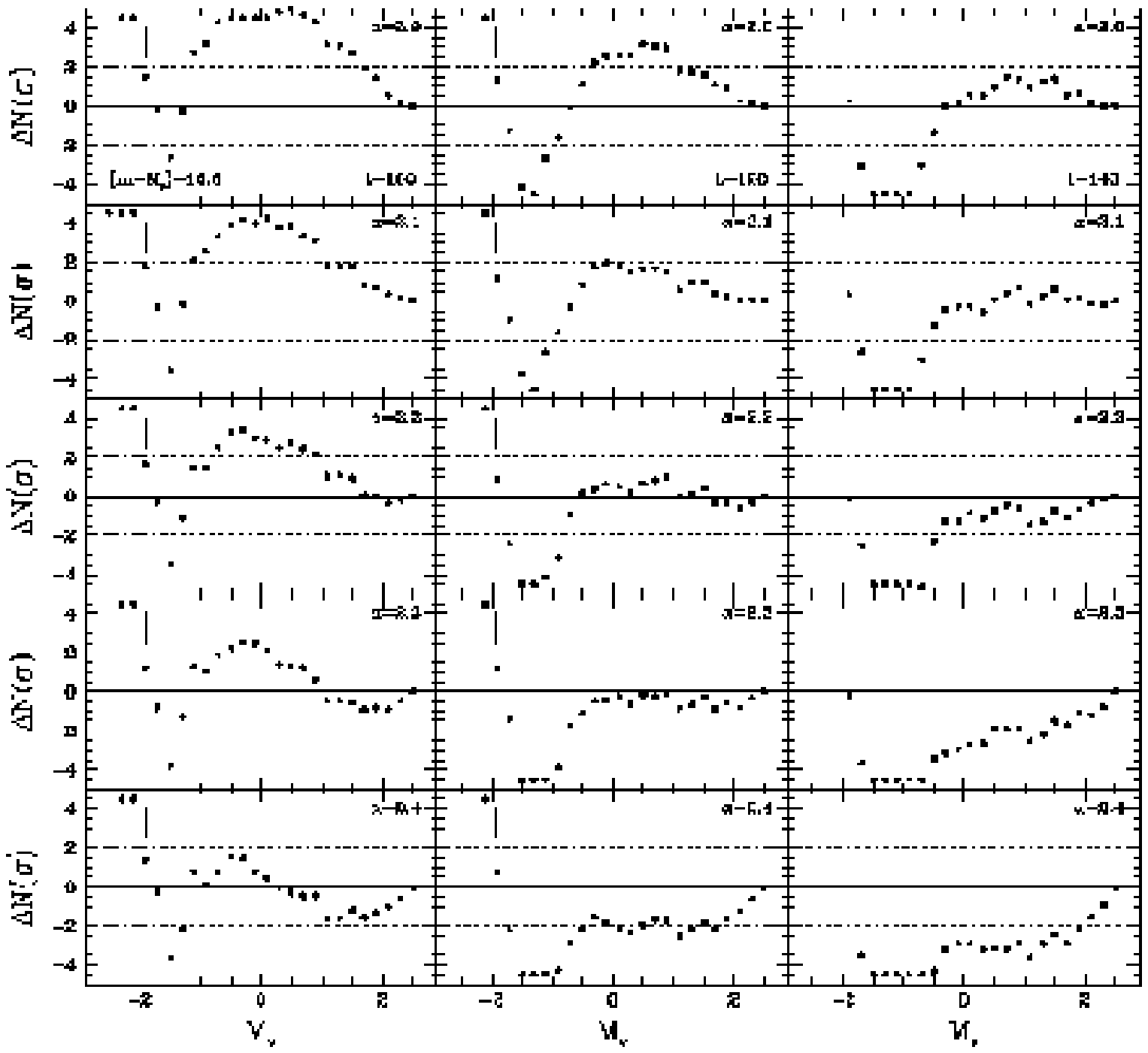]{Comparison between the theoretical MSILF 
 and the observed data  in the standard case, in units of
 $\sigma=(N_{obs}-N_{theo})/(N_{theo})^{1/2}$. Rows and columns are 
as in Figure \ref{fig23}.\label{fig24}}

\figcaption[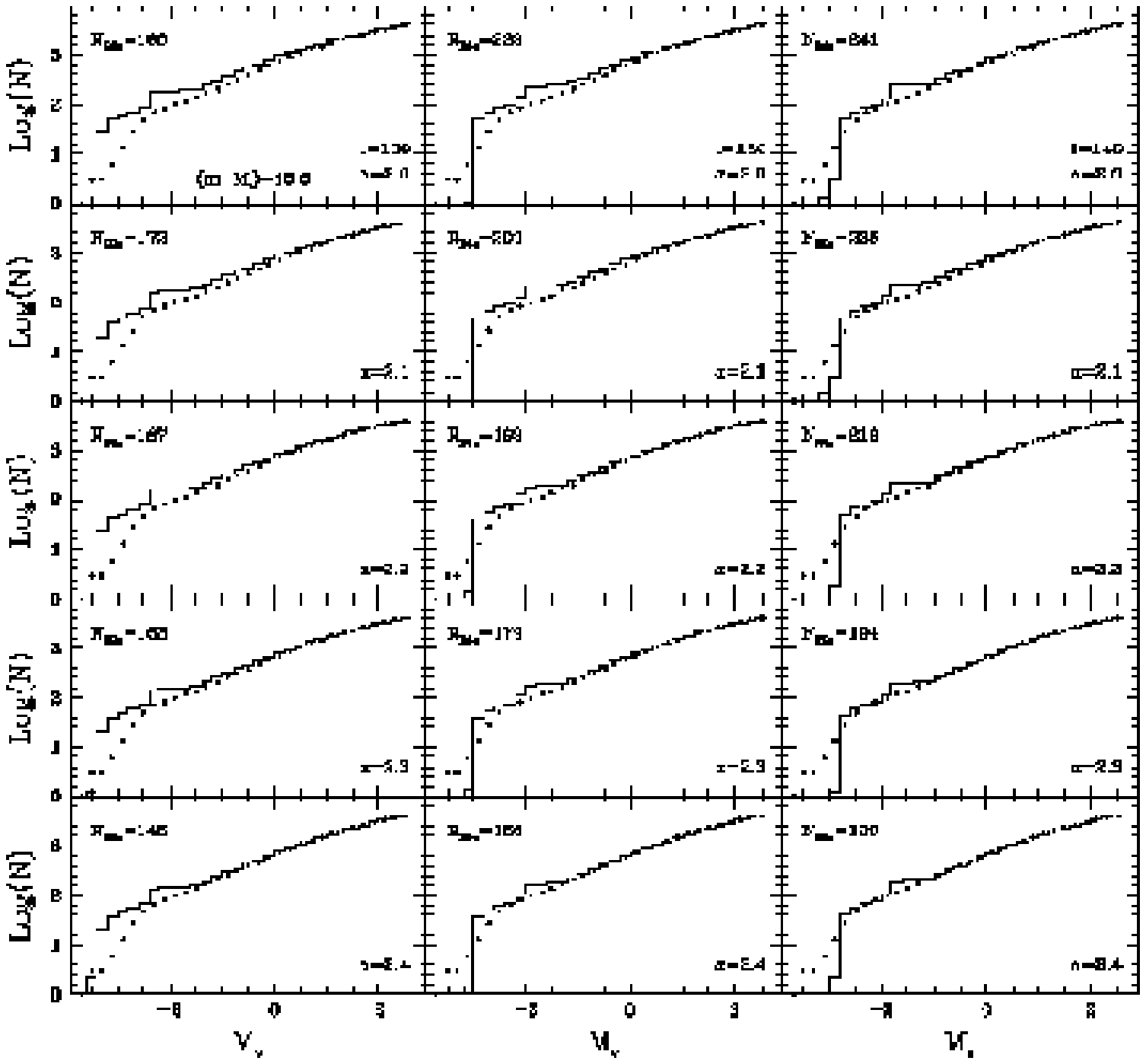]{Comparison of the integrated theoretical and 
observed LF including the giant stars. Symbols and arrangement of rows
and columns are as in Figure \ref{fig23}.\label{fig25}}

\figcaption[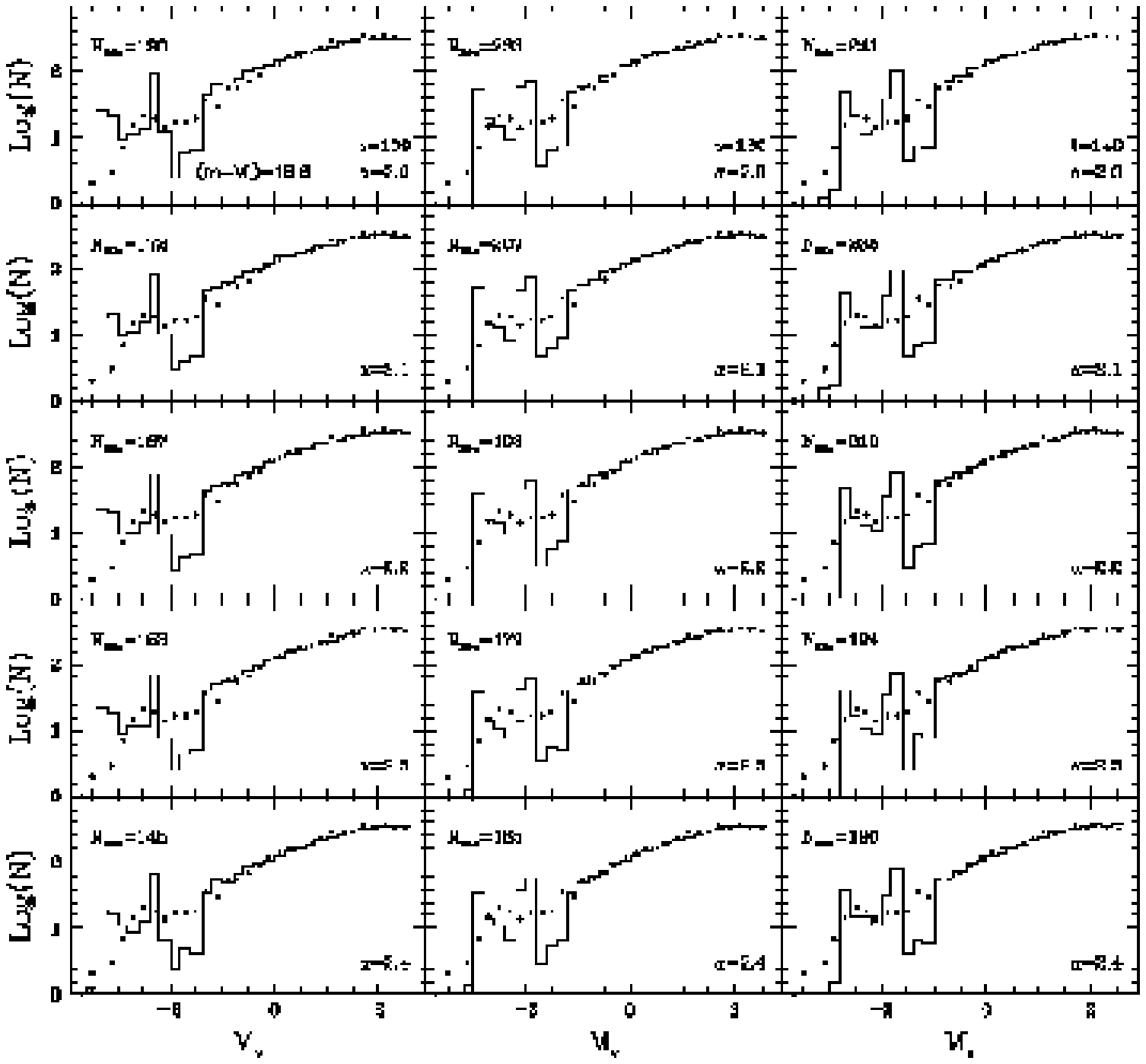]{As in Figure \ref{fig25} but for the differential
 LF. \label{fig26}}

\figcaption[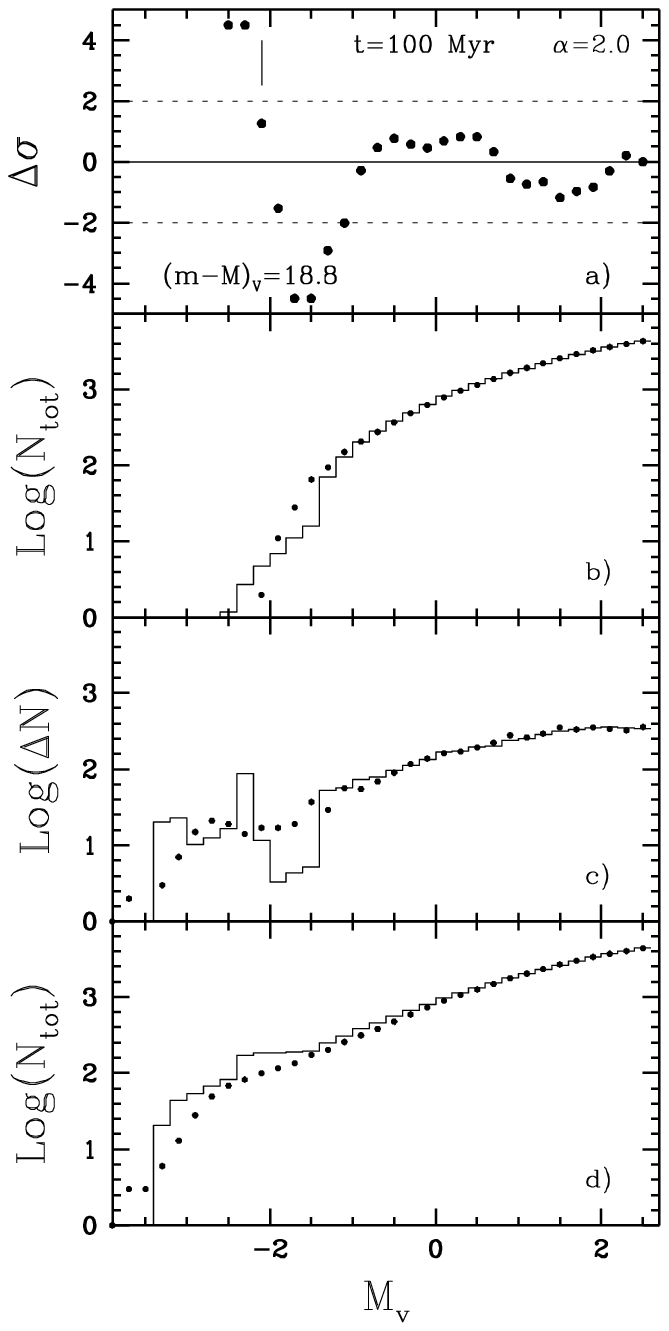]{The best fit obtained in the standard case for 
the ``long'' distance modulus $(m-M)_V=18.8$.\label{fig27}}

\figcaption[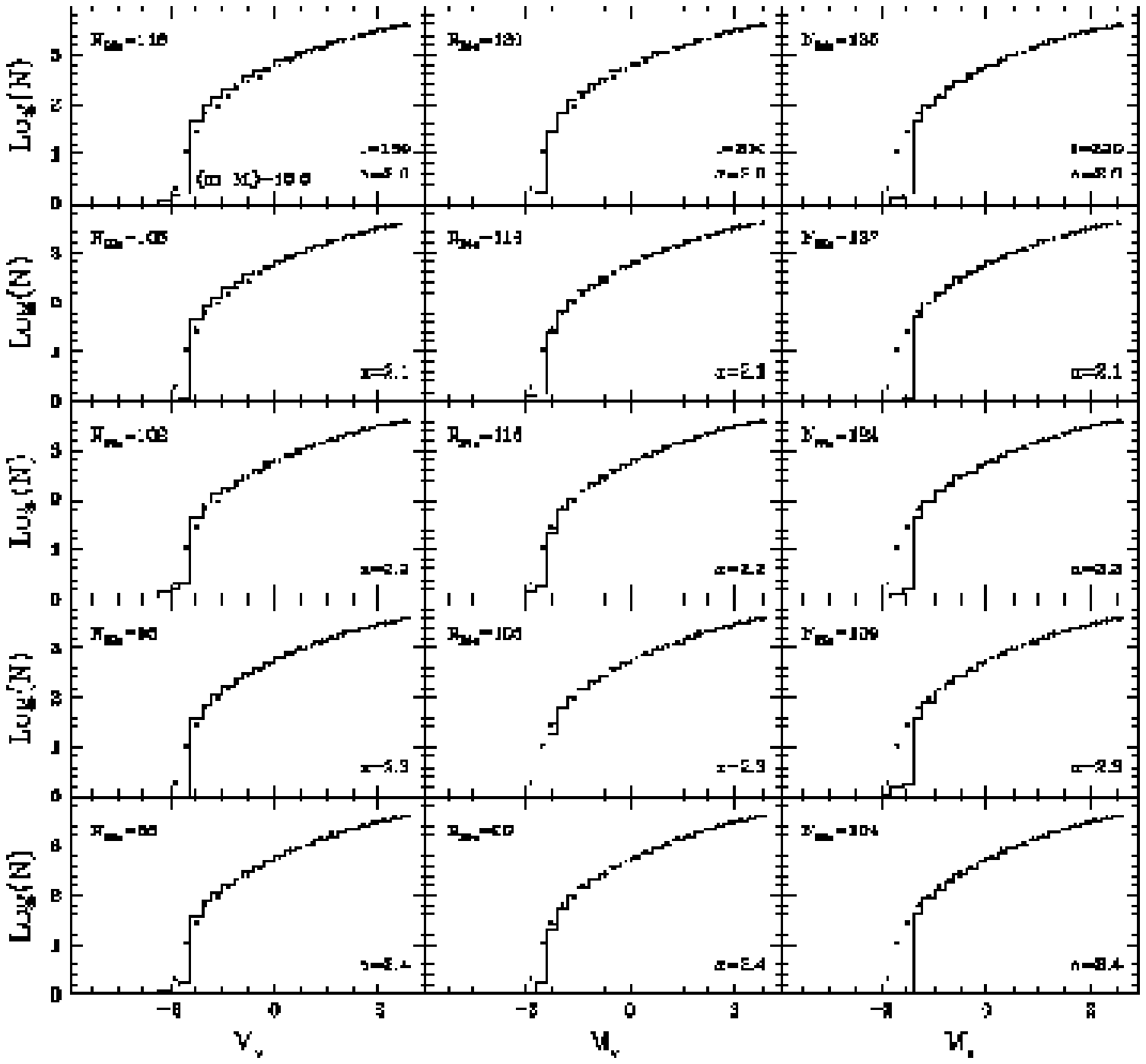]{Comparison between the theoretical MSILF 
(histogram) and the observed data (dots) in the AECC case for three ages
(columns) and five values of $\alpha$ (rows) and $(m-M)_V=18.6$.\label{fig28}}

\newpage
\figcaption[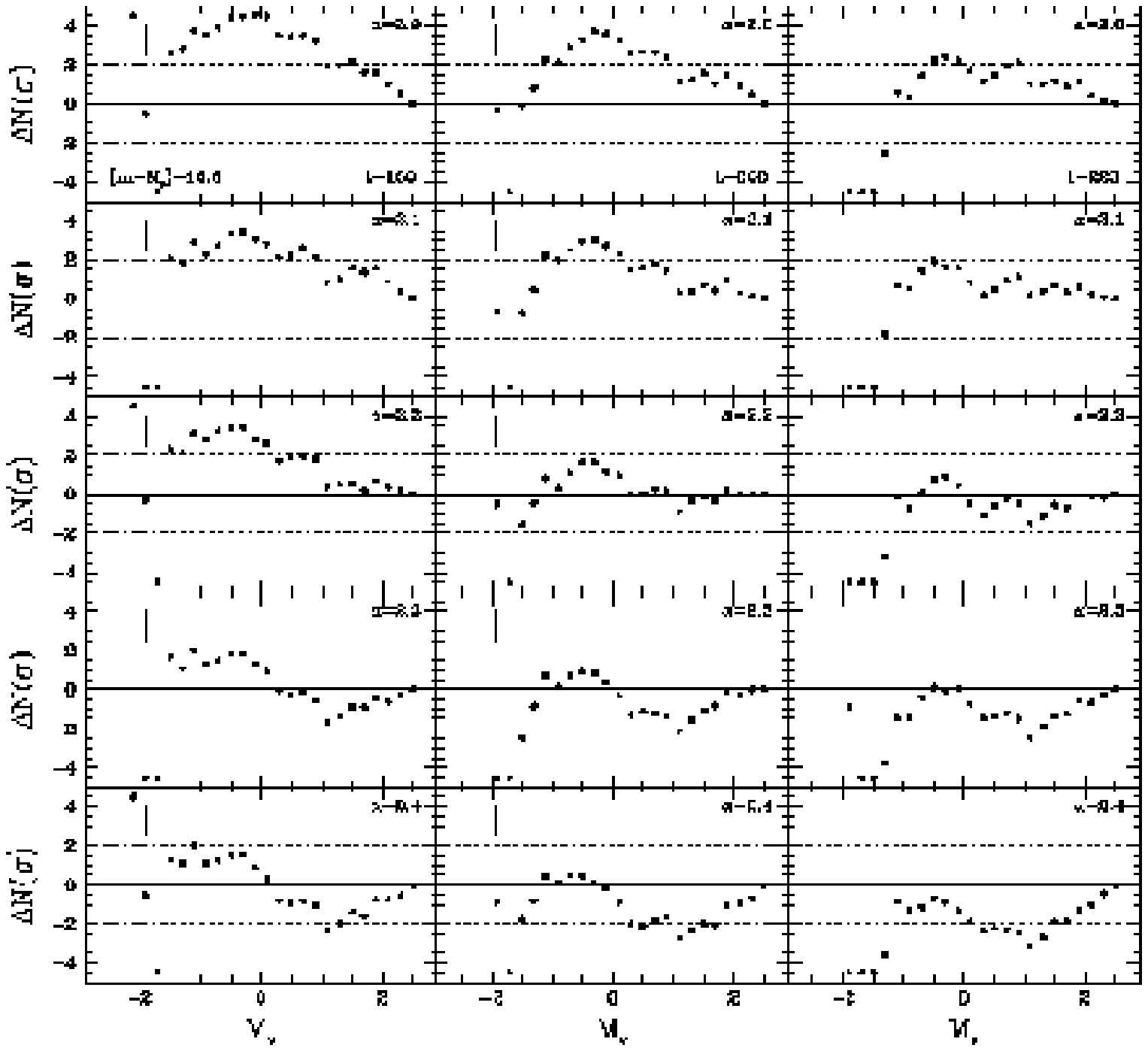]{Comparison between the theoretical MSILF 
 and the observed data  in the AECC case, in units of
 $\sigma=(N_{obs}-N_{theo})/(N_{theo})^{1/2}$. Rows and columns are 
 as in Figure \ref{fig28}.\label{fig29}}

\figcaption[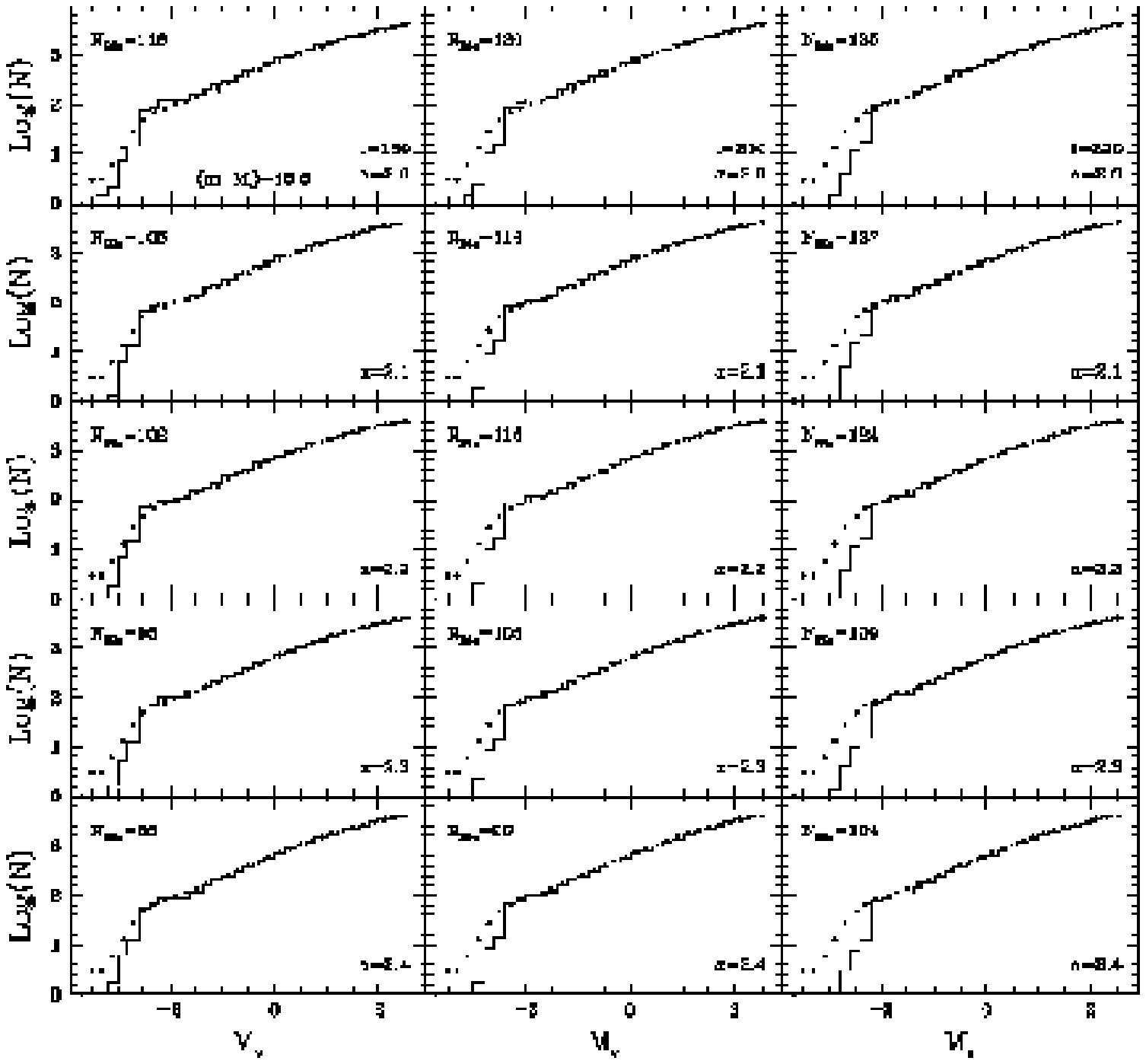]{Comparison of the integrated theoretical and 
observed LF including the giant stars for the AECC case. 
Symbols and arrangement of rows and columns are as in Figure 
\ref{fig28}.\label{fig30}}

\figcaption[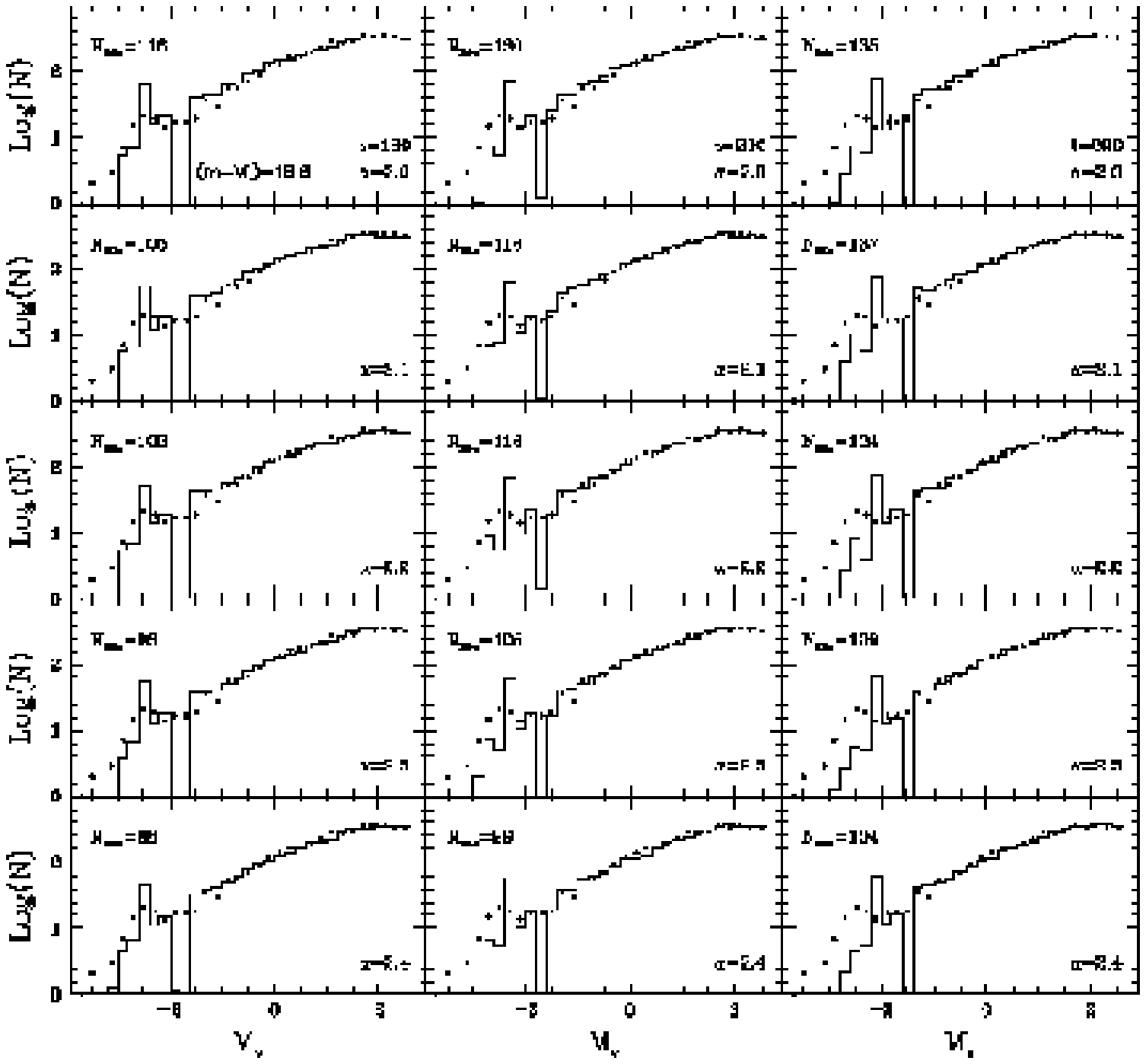]{As in Figure \ref{fig30} but for the differential
 LF. \label{fig31}}

\figcaption[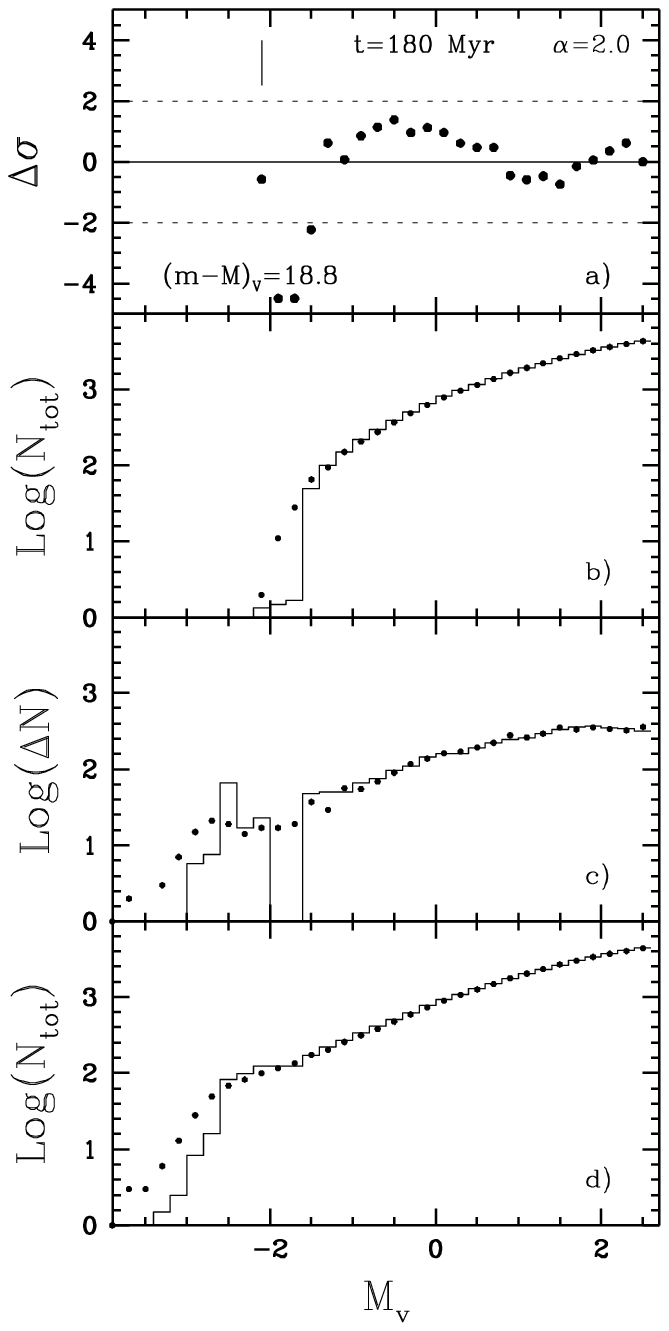]{The best fit for the AECC case for the ``long''
distance modulus $(m-M)_V=18.8$.\label{fig32}}

\figcaption[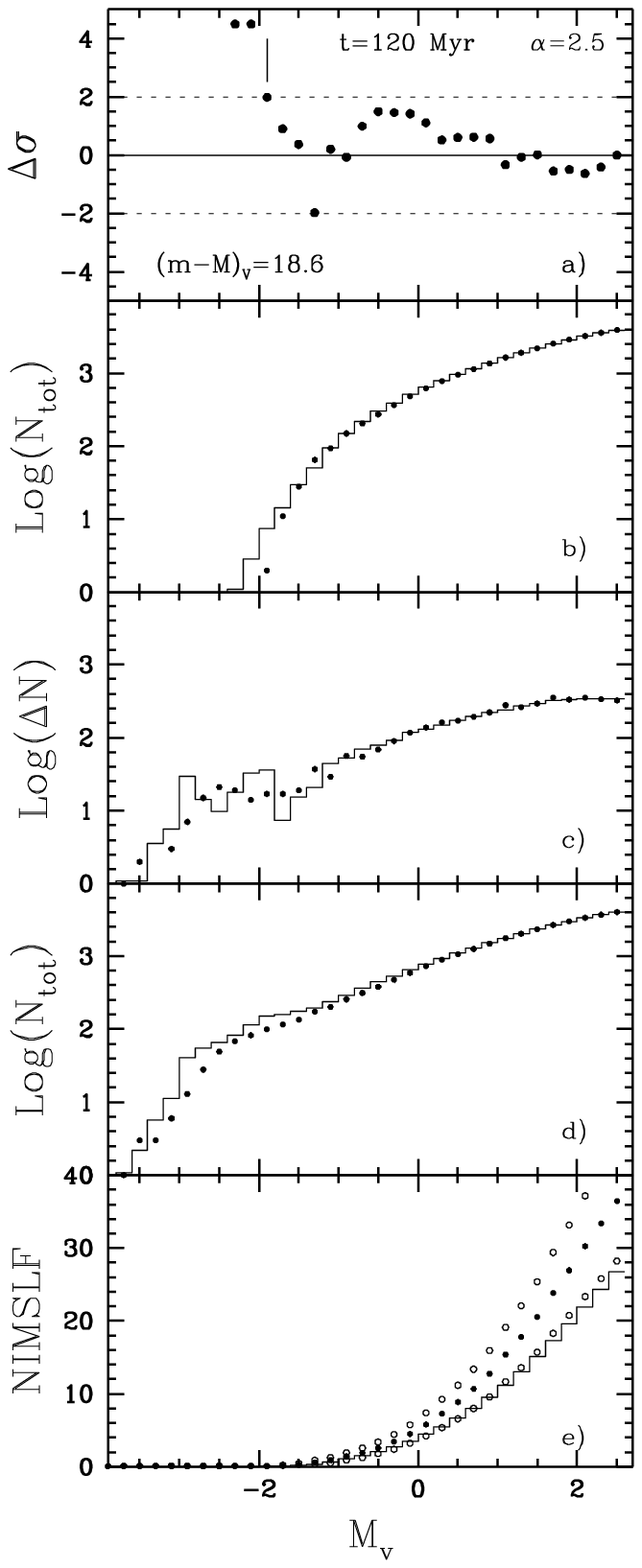]{The best fit obtained for the standard+binaries
case for the ``short'' distance modulus $(m-M)_V=18.6$.\label{fig33}}

\figcaption[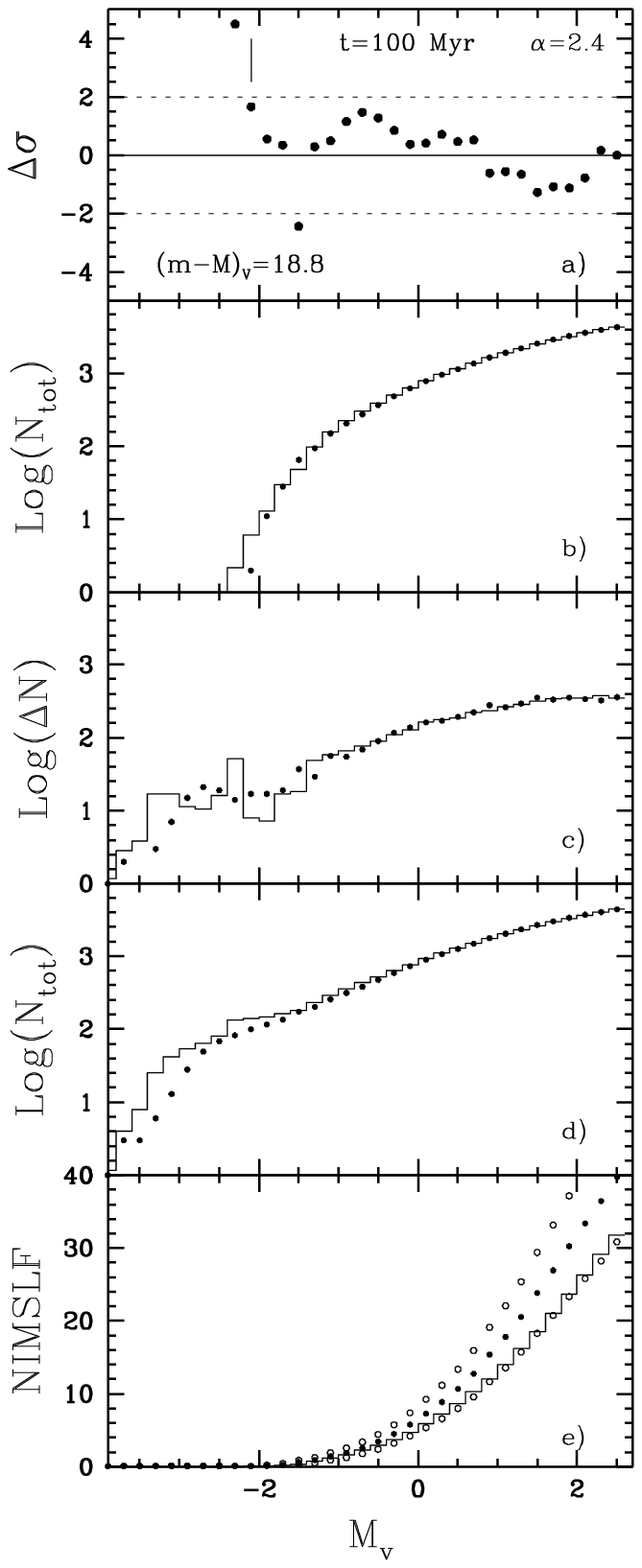]{As in Figure \ref{fig33} but for the ``long''
distance modulus $(m-M)_V=18.8$.\label{fig34}}

\figcaption[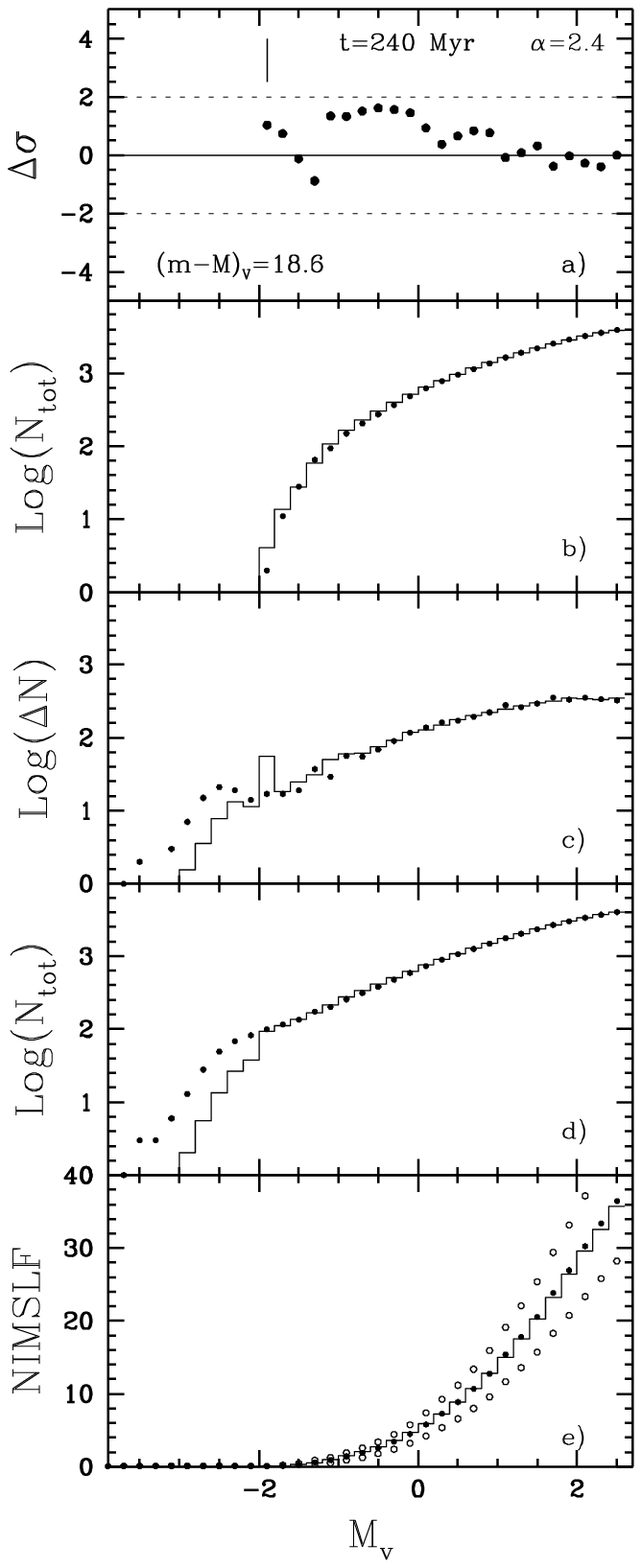]{As in Figure \ref{fig33} but for the AECC+binaries
case.\label{fig35}}

\end{document}